\documentclass[journal,10pt]{IEEEtran} 

\usepackage{amsmath}
\usepackage{amssymb}

\usepackage{pgf,tikz}
\usepackage{pgfplots}
\pgfplotsset{compat=1.5}
\usepackage{adjustbox}
\usepackage{balance}
\usepackage{newtxtext} 
\usepackage{mathptmx} 
\usepackage{mathtools}
\usepackage{mathrsfs}
\usepackage{xcolor}
\usepackage{caption}
\usepackage{subcaption}
\usepackage{cite}
\usepackage{graphicx}
\usepackage[english]{babel}
\usepackage{floatrow}
\makeatletter
\newcommand{\mathleft}{\@fleqntrue\@mathmargin8pt}
\newcommand{\mathcenter}{\@fleqnfalse}
\makeatother

\newcommand{\Figref}[1]{Fig.~\ref{#1}}

\DeclarePairedDelimiter{\ceil}{\lceil}{\rceil}

\newtheorem{proposition}{Proposition}
\newtheorem{remark}{Remark}

\begin{document}
\title{ Rapid Node Cardinality Estimation in Heterogeneous Machine-to-Machine Networks}
\author{Sachin Kadam, Sesha Vivek Y., P. Hari Prasad, Rajesh Kumar, and Gaurav S. Kasbekar}
\maketitle{}
{\renewcommand{\thefootnote}{} \footnotetext{S. Kadam and G. Kasbekar are with Department of Electrical Engineering, Indian Institute of Technology (IIT) Bombay, Mumbai, India. Sesha Vivek Y. is with Goldman Sachs, Bengaluru, India, P. Hari Prasad is with Daikin Industries Limited, Osaka, Japan, and R. Kumar is with Integrated Test Range, DRDO, Balasore, India. Their email addresses are sachink@ee.iitb.ac.in, gskasbekar@ee.iitb.ac.in, seshavivek.yenduri@gs.com, hari.prasad@daikin.co.jp, and rajesh.kumar@itr.drdo.in respectively. Sesha Vivek Y., P. Hari Prasad, and R. Kumar worked on this research while they were with IIT Bombay. The contributions of S. Kadam and G. Kasbekar have been supported by SERB grant SB/S3/EECE/157/2016.
\par A preliminary version of this paper appeared in Proc. of IEEE VTC2019-Spring Decentralized Technologies and Applications for IoT (D'IoT)~\cite{Kada1904:Rapid}. 
}}

\begin{abstract}	
Machine-to-Machine (M2M) networks are an emerging technology with applications in various fields, including smart grids, healthcare, vehicular telematics and smart cities. Heterogeneous M2M networks contain different types of nodes, e.g., nodes that send emergency, periodic, and normal type data. An important problem is to rapidly estimate the number of active nodes of each node type in every time frame in such a network. In this paper, we design two schemes for estimating the active node cardinalities of each node type in a heterogeneous M2M network with $T$ types of nodes, where $T \ge 2$ is an arbitrary integer. Our schemes consist of two phases--  in phase 1, coarse estimates are computed, and in phase 2, these estimates are used to compute the final estimates to the required accuracy. We analytically derive a condition for one of our schemes that can be used to decide as to which of two possible approaches should be used in phase 2 to minimize its execution time. The expected number of time slots required to execute and the expected energy consumption of each active node under one of our schemes are analysed. Using simulations, we show that our proposed schemes require significantly fewer time slots to execute compared to estimation schemes designed for a heterogeneous M2M network in prior work, and also, compared to separately executing a well-known estimation protocol designed for a homogeneous network in prior work $T$ times to estimate the cardinalities of the $T$ node types, even though all these schemes obtain estimates with the same accuracy.
\end{abstract}

\section{Introduction}\label{Intro}
Machine-to-Machine (M2M) communications is emerging  as a key technology for connecting together a large number of autonomous devices that require minimal to zero human intervention in order to generate, process, and transmit data~\cite{wu2011m2m}. M2M networks have extensive applications in various fields including  smart grids, health care, vehicular telematics, smart cities, security and public safety,   agriculture, and industrial automation~\cite{liu2014design}. 

The problem of designing efficient networking protocols to cater to the increasing number of M2M devices is an active research area~\cite{liu2014design}. 
In particular, the design of medium access control (MAC) protocols for M2M networks is challenging because they have a number of unique characteristics, e.g., (i) network access needs to be provided to an extremely large number of M2M devices, (ii) most M2M devices are battery powered and have limited power availability, (iii) the quality  of service (QoS) requirements in M2M applications differ from those in Human-to-Human (H2H) communications and are also different for different M2M devices~\cite{rajandekar2015survey}. A key component of a MAC protocol for M2M networks is an estimation protocol that rapidly estimates the number of active devices (i.e., the devices that currently have some data that needs to be sent to the base station) in every time frame~\cite{rajandekar2015survey}. These estimates can be used to find the optimal values of various parameters of the MAC protocol, e.g., contention probability, contention period, data transmission period etc, in each time frame~\cite{M2MTanab, M2MDuan, Est_CSMA, Congestion_Liu, M2MOh, M2MMort, TCP_Bac, DQ_Bui, LTE_Lin, Raptor_Shir}. For example, recall that for the Slotted ALOHA protocol, the optimal contention probability is the reciprocal of the number of active nodes~\cite{bertsekas1992data}.   

There has been extensive research on the problem of node cardinality estimation in M2M networks and in Radio Frequency Identification (RFID) systems (see Section~\ref{Rel_Work} for a review of these papers); however, with the exception of our prior work~\cite{kadam2017fast},~\cite{TechReport2018},\footnote{Note that~\cite{TechReport2018} is an extended version of the conference paper~\cite{kadam2017fast}.} all the papers in the existing research literature address the problem of node cardinality estimation in a \emph{homogeneous} network, i.e., a network consisting of only one type of nodes. In contrast, in this paper, we address the problem of obtaining separate estimates of the number of active nodes of each type in a \emph{heterogeneous} network, i.e., a network with multiple types of nodes. Note that executing a node cardinality estimation protocol for a homogeneous network multiple times to obtain the active node cardinalities of each type in a heterogeneous network is inefficient. 
In this paper, we consider an M2M network containing $T$  types of nodes, where $T \ge 2$ is an arbitrary integer, which we refer to as Type 1 ($\mathscr{T}_1$), \ldots, Type $T$ ($\mathscr{T}_T$) nodes; e.g., these may be emergency, periodic, normal data type nodes etc. 
We design two estimation schemes to rapidly obtain separate estimates of the number of active nodes of each data type in a heterogeneous M2M network with $T$ types of nodes. Both these schemes outperform the schemes proposed in our prior work~\cite{kadam2017fast},~\cite{TechReport2018} (see Section~\ref{Rel_Work} for details). 

The main contributions of this paper are as follows. 
\begin{itemize}
\item We propose two schemes, viz., the heterogeneous SRC$_S$-1 scheme (HSRC-1) and  the heterogeneous SRC$_S$-2 scheme (HSRC-2), for rapid node cardinality estimation in heterogeneous networks by extending the simple RFID counting (SRC$_S$) protocol proposed  for a homogeneous network in~\cite{zhou2016understanding}. 
\item Our proposed schemes consist of two phases and one of two possible approaches is used in phase 2. We analytically derive a condition, which can be used to find out as to which approach should be used in phase 2 of HSRC-1 in order to minimize its execution time.  Also, we validate this condition via simulations. 
\item We mathematically analyze the expected number of time slots required by HSRC-1 to execute and the expected energy consumption of a node under the scheme. 
\item We evaluate the performances of both the proposed estimation schemes, HSRC-1 and HSRC-2, via extensive simulations and show that they require significantly fewer time slots to execute than the estimation scheme in which the SRC$_S$ protocol is separately executed $T$ times to estimate the cardinalities of the $T$ node types,  as well as the  estimation schemes proposed in~\cite{kadam2017fast},~\cite{TechReport2018}, even though all these schemes obtain estimates with the same accuracy.  
\end{itemize}

The rest of this paper is organized as follows. A review of related prior literature is provided in Section~\ref{Rel_Work}. The network model and problem formulation are described and relevant background is reviewed in Section~\ref{SC:nw:model:prb:form:background}. The rapid node cardinality estimation schemes for heterogeneous M2M networks proposed in this paper, HSRC-1 and  HSRC-2, are described in Section~\ref{EstScheme}.  A condition that can be used to find out as to which of two possible approaches should be used in phase 2 of HSRC-1 in order to minimize its execution time is analytically derived in Section~\ref{SC:HSRC1:phase2:condition}. The expected number of time slots required by HSRC-1 to execute and the expected energy consumption of a node under the scheme are mathematically analysed in Section~\ref{Analysis}. We evaluate the performances of our proposed estimation schemes via simulations in Section~\ref{Simu}. Finally, we provide conclusions in Section~\ref{Conc}.  

\section{Related Work}\label{Rel_Work}
Owing to the importance of active node cardinality estimation as part of the design of a MAC protocol, extensive research has been carried out on the problem of estimating the number of active devices in a homogeneous M2M network~\cite{M2MTanab, M2MDuan, Est_CSMA, Congestion_Liu, M2MOh, M2MMort, TCP_Bac, DQ_Bui, LTE_Lin, Raptor_Shir}. Also, in~\cite{M2MTanab, M2MDuan, Est_CSMA, Congestion_Liu, M2MOh, M2MMort, TCP_Bac, DQ_Bui, LTE_Lin, Raptor_Shir}, using the estimates obtained, the contention probabilities that maximize the throughput of their respective MAC protocols for M2M networks are determined.    
In~\cite{M2MDuan}, the proposed estimation scheme uses the estimates computed in the previous frame and the sub-optimal Dynamic Access Class Barring (D-ACB) factors of the previous frame to estimate the number of active nodes present in the current frame.
In~\cite{Est_CSMA},  a modified version of the CSMA/CA protocol is proposed for an M2M network, which uses the size of the preceding backoff window and previously computed active node cardinality estimates to compute the size of the backoff window to be used in the current frame.
In~\cite{M2MDuan, Est_CSMA}, the estimates used in the current frame are computed using the estimates obtained in previous frames, whereas in our work, the estimates of different frames are independently computed.
In~\cite{M2MOh}, a new scheme for dynamic access control and random access channel resource allocation  based on an estimation scheme is proposed. The estimation scheme used in~\cite{M2MOh} uses only the number of idle slots to compute estimates, whereas our work uses the number of idle slots as well as the numbers of slots in which successful transmissions and collisions take place.
In~\cite{Congestion_Liu}, a novel 6-Dimensional Markov Chain (6-DMC) based estimation scheme to estimate the number of delay tolerant devices (DTDs) and delay sensitive devices (DSDs) is proposed. 
The estimation scheme in~\cite{Congestion_Liu} (respectively,~\cite{M2MMort},~\cite{M2MTanab}) uses the 6-DMC (respectively, Maximum Likelihood Estimation (MLE), M2M-OSA, an extension of the opportunistic splitting algorithm (OSA)) based estimation scheme, whereas in our work, we use the SRC$_S$ based estimation scheme~\cite{zhou2016understanding}.
 A  satellite random access (RA) MAC protocol is proposed in~\cite{TCP_Bac}, wherein an estimate of the number of Return Channel Satellite Terminals (RCSTs) is computed and used in throughput maximization. The length of the current frame in the model in~\cite{TCP_Bac} depends on the number of collisions in the previous frames, whereas in our model, the length of each frame is fixed and constant.
In the scheme proposed in~\cite{DQ_Bui}, the number of nodes that cause collisions is estimated  so that nodes can be efficiently divided into a fixed number of groups such that intra-group collisions are minimized, thus improving the throughput in Long-Term Evolution (LTE) networks. In~\cite{DQ_Bui}, cardinality estimation of only the nodes that cause collisions is performed, whereas our proposed schemes estimate the cardinalities of all active nodes.
A novel channel contention resolution scheme, viz., Dynamic Backoff (DB), is proposed in~\cite{LTE_Lin}, which  estimates the number of active devices that attempt to contend to send preambles; the size of the backoff window used to contend on the channel for data transfer is adjusted using the computed estimate. The size of each frame is dynamically adjusted in the scheme proposed in~\cite{LTE_Lin} based on the estimated number of devices, whereas in our model, the size of each frame is fixed.
A load estimation algorithm is proposed in~\cite{Raptor_Shir}, in which the base station (BS) detects preambles and estimates the number of active devices using the history of transmissions that have selected each preamble. In~\cite{Raptor_Shir}, node cardinality estimates are computed using the history of transmissions, whereas in our work, they are computed using only transmissions in the current frame.

The problem of node cardinality estimation in M2M networks is similar to that of tag cardinality estimation in the context of RFID technology. In particular, in the latter context, an RFID reader estimates the number of tags, similar to the former context, in which a base station estimates the number of active nodes in an M2M network. Schemes for estimating the number of tags in an RFID system have been proposed in~\cite{kodialam2007anonymous, qian2011cardinality, zheng2012pet,zheng2013zoe,gong2014arbitrarily,zhou2016understanding, zhou2018counting, RFIDLiu1, RFIDLiu2, RFIDLiu3, RFIDChen1, RFIDChen2, RFIDGong1, RFIDGong2, RFIDGong3}. 

However, all of the above node cardinality estimation schemes~\cite{M2MTanab, M2MDuan, Est_CSMA, Congestion_Liu, M2MOh, M2MMort, TCP_Bac, DQ_Bui, LTE_Lin, Raptor_Shir},~\cite{kodialam2007anonymous, qian2011cardinality, zheng2012pet,zheng2013zoe,gong2014arbitrarily, zhou2016understanding, zhou2018counting, RFIDLiu1, RFIDLiu2, RFIDLiu3, RFIDChen1, RFIDChen2, RFIDGong1, RFIDGong2, RFIDGong3} are designed for node cardinality estimation in \emph{homogeneous} networks. In contrast, in this paper, we propose node cardinality estimation schemes for \emph{heterogeneous} networks with $T$ types of nodes, where $T \geq 2$ is an arbitrary integer.  

Now, after carefully reviewing various estimation protocols, including Enhanced Zero-Based estimator~\cite{kodialam2007anonymous}, Lottery Frame (LoF) based estimator~\cite{qian2011cardinality}, Probabilistic Estimating Tree estimator~\cite{zheng2012pet}, Zero-One estimator~\cite{zheng2013zoe}, and Arbitrarily Accurate Approximation estimator~\cite{gong2014arbitrarily}, the authors of~\cite{zhou2016understanding} have shown that for an estimation protocol for a homogeneous network to be efficient, \emph{i.e.}, for it to take the minimum possible number of time slots to estimate the node cardinality for a given set of accuracy specifications, it is necessary that the protocol  have two phases-- a phase for obtaining a coarse estimate, followed by a phase that uses the coarse estimate to achieve an accuracy target. Also, the authors of~\cite{zhou2016understanding} have  devised an improved protocol, viz., the simple RFID counting (SRC$_S$) protocol, which has two phases, for tag cardinality estimation in homogeneous RFID networks. In this paper, we propose two schemes for rapid node cardinality estimation in heterogeneous networks by extending the SRC$_S$ protocol proposed  for a homogeneous network in~\cite{zhou2016understanding}. Both the proposed schemes have two phases, which correspond to the two phases in the SRC$_S$ protocol. 

To the best of our knowledge, in prior literature there is only one work, viz., our prior work~\cite{kadam2017fast},~\cite{TechReport2018}, which designs node cardinality estimation schemes for heterogeneous M2M networks. 
We have shown in this paper, via simulations, that the estimation schemes proposed in this paper significantly outperform those in~\cite{kadam2017fast},~\cite{TechReport2018} in terms of the number of time slots required to execute for achieving a given level of estimation accuracy. Intuitively, this is because the former (respectively, latter) are designed by extending the SRC$_S$ protocol~\cite{zhou2016understanding} (respectively, LoF based protocol~\cite{qian2011cardinality})\footnote{Note that both the SRC$_S$ protocol~\cite{zhou2016understanding} and LoF based protocol~\cite{qian2011cardinality} are estimation protocols for a \emph{homogeneous} network.} for node cardinality estimation in a heterogeneous network, and the SRC$_S$ protocol~\cite{zhou2016understanding}  has been shown to outperform the LoF based protocol~\cite{qian2011cardinality} in~\cite{zhou2016understanding} in terms of the number of time slots required to execute.

\section{Network Model, Problem Formulation and Background}
\label{SC:nw:model:prb:form:background}
\subsection{The Node Cardinality Estimation Problem in a Heterogeneous M2M Network}\label{nwmodel}
\begin{figure}
	\begin{center}
		\includegraphics[scale = 0.4]{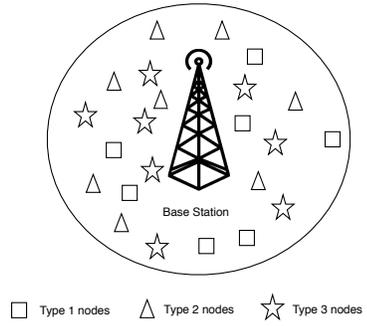}
		\caption{A base station with $T = 3$ different types of nodes within its range.}
		\label{NetworkModel}
	\end{center}
\end{figure}

Consider a heterogeneous M2M network consisting of a base station (BS) and $T$ different types-- say Type 1 ($\mathscr{T}_1$), \ldots, Type T ($\mathscr{T}_T$)-- of nodes within its range,  where $T \ge 2$ is an arbitrary integer. Fig.~\ref{NetworkModel} illustrates such a network for the case $T=3$. Time is divided into frames of equal durations, and in each frame only a subset of the nodes of each type are \emph{active}, i.e., have data to send to the BS.  Let $n_b$  be the number of active nodes  of Type $b$, $b \in \{1, \ldots, T\}$, in a given frame. Our objective is to rapidly estimate the values of $n_b, \ b \in \{1,\ldots, T\}$. 

In particular, let $\hat{n}_b$ be the estimated value of $n_b$. Let $\delta$, the desired error probability, and $\epsilon$, the desired relative error bound, be the user specified accuracy requirements, i.e., the parameters with which the estimate $\hat{n}_b$ needs to be obtained.  Our objective is to rapidly find estimates $\hat{n}_b$ for $n_b$, $b \in \{1,\ldots, T\}$, such that $P(| \hat{n}_b - n_b|$ $\leq$ $\epsilon n_b)$ $\geq 1 - \delta$, $\forall b \in \{1,\ldots, T\}$. Note that we assume that the accuracy requirement parameters $\epsilon$ and $\delta$ are the same for all the $T$ node types.

\begin{figure}
\begin{center}
	\includegraphics[scale = 0.7]{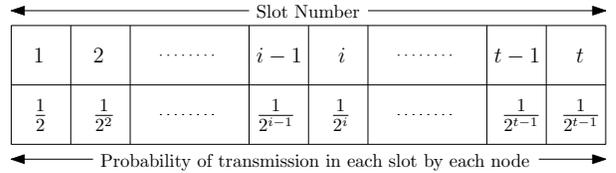}
	\caption{The figure shows a single trial of the LoF based protocol.}
	\label{LoF_window}
\end{center}
\end{figure}

\subsection{Review of Lottery Frame ($LoF$) based Protocol~\cite{qian2011cardinality}}\label{SubSec_LoF}
Our proposed schemes extend the Simple RFID Counting (SRC$_S$) protocol, which was proposed in~\cite{zhou2016understanding} for node cardinality estimation in a homogeneous network, for node cardinality estimation in a heterogeneous M2M network with $T$ types of nodes. The SRC$_S$ protocol consists of two phases and in phase 1, it uses the LoF based protocol, which was designed in~\cite{qian2011cardinality} and uses the probabilistic bitmap counting technique proposed in~\cite{flajolet1985probabilistic}, for node cardinality estimation in homogeneous networks. So we provide a brief review of the LoF based protocol (respectively, the SRC$_S$ protocol) in this subsection (respectively, in Section~\ref{SubSec_Est}).

The LoF based protocol is designed for finding an estimate, say $\hat{n}$, of the number of active nodes, say $n$, in a homogeneous network to within given accuracy requirements $\epsilon$ and $\delta$. That is, the user requires that $P(| \hat{n} - n|$ $\leq$ $\epsilon n)$ $\geq 1 - \delta$.
Let $n_{all}$ be the total number of nodes manufactured and $t = \ceil{\log_{2}{n_{all}}}$.\footnote{$\ceil{x}$ denotes the smallest integer greater than or equal to $x$.} 

The LoF based protocol consists of multiple independent trials, each consisting of $t$ time slots. Let $M$ be the minimum number of trials required by the LoF based protocol to obtain an estimate of $n$ to within the given accuracy requirements $\epsilon$ and $\delta$. $M$ is given by the following expression~\cite{qian2011cardinality}:
\begin{equation*}
M = \ceil[\Bigg]{\max\left( \left[ \frac{-1.1213c}{\log_2 (1-\epsilon)}\right]^2, \left[ \frac{1.1213c}{\log_2 (1+\epsilon)}\right]^2 \right)},
\end{equation*}
where $c = \sqrt{2} \times erf^{-1} (1 - \delta)$ and $erf^{-1} (\cdot)$ is the inverse Gaussian error function.

Fig.~\ref{LoF_window} shows a single trial of the LoF based protocol. In the $m^{th}$ trial, $m \in \{1, \ldots, M\}$, every active node randomly chooses the $i^{th}$ slot with probability:
\begin{equation}
\label{EQ:LoF:bar:pi}
\bar{p}(i) = \left\{ 
\begin{array}{ll}
1/2^{i}, & \mbox{for } i \in \{1, \ldots, t-1\}, \\
1/2^{t-1}, & \mbox{for } i = t. \\
\end{array}
\right. 
\end{equation}
Each active node transmits in its chosen slot. After the trial, each slot of the trial can be in one of the following three states: (i) \emph{Empty}: No node  transmitted in that slot, (ii) \emph{Success}: Exactly one node transmitted in that slot, (iii) \emph{Collision}: More than one node transmitted in that slot. Let $j(m)$ be the smallest number $j \in \{1, \ldots, t\}$, such that the $j^{th}$ slot is in the \emph{Empty} state in the $m^{th}$ trial.\footnote{If none of the $t$ slots are in the \emph{Empty} state in the $m^{th}$ trial, then $j(m) = t$.}  At the end of all $M$ trials, the estimate of $n$ is computed as~\cite{qian2011cardinality}:
\begin{equation}
\label{EQ:LoF:estimate:nhat}
\hat{n}=1.2897 \times 2^{\Sigma_{m=1}^M (j(m) - 1)/M}.
\end{equation}

\subsection{Review of Simple RFID Counting (SRC$_S$) Protocol~\cite{zhou2016understanding}}\label{SubSec_Est}
We now review the SRC$_S$ protocol, which is a  protocol designed in~\cite{zhou2016understanding} for node cardinality estimation to within given accuracy requirements, $\epsilon$ and $\delta$, in homogeneous networks, and which we extend for node cardinality estimation in heterogeneous networks.  

\begin{figure}
\begin{center}
	\includegraphics[scale = 0.48]{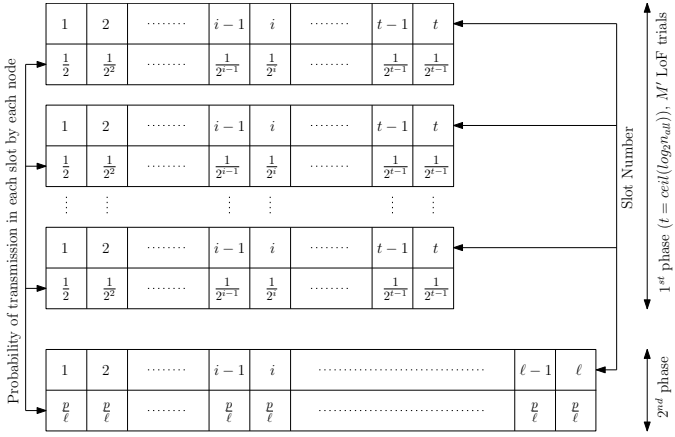}
	\caption{The figure shows the frame structure used in the SRC$_S$ protocol.}
	\label{SRCs_window}
\end{center}
\end{figure}

Let the number of active nodes in a given homogeneous network be $n$. The SRC$_S$ protocol is a two phase protocol (see Fig.~\ref{SRCs_window}); at the end of phase 1 (respectively, phase 2), it finds a rough estimate $\tilde{n}$ (respectively, the final estimate $\hat{n}$) of $n$~\cite{zhou2016understanding}. Phase 1 (respectively, phase 2) of the protocol consists  of a sequence of trials (respectively, a single trial), and each trial consists of multiple  slots. The number of slots in a trial is called the length of the trial. After a trial, a slot of the trial can be in one of the following three states: \emph{Empty}, \emph{Success} or \emph{Collision}. These states have the same meanings as in Section~\ref{SubSec_LoF}.

Phase 1 of the SRC$_S$ protocol consists of a sequence of independent trials of the LoF based protocol~\cite{qian2011cardinality} (see Section~\ref{SubSec_LoF} for a review); let $M^{\prime}$ be the number of trials of the LoF based protocol conducted. For each $m \in \{1, \ldots, M^{\prime}\}$, let $j(m)$ be as defined in Section~\ref{SubSec_LoF}. At the end of all $M^{\prime}$ trials, the rough estimate of $n$ is computed as  $\tilde{n}=1.2897 \times 2^{\Sigma_{m=1}^{M^{\prime}} (j(m) - 1)/M^{\prime}}$ (see \eqref{EQ:LoF:estimate:nhat}).  The number of trials, $M^{\prime}$, is determined based on the desired error probability $\delta$. For example, for $\delta=0.2$, $M^{\prime} = 10$ is used~\cite{zhou2016understanding}.  
 
Let $BB$ denote the ``balls-and-bins'' method~\cite{zhou2016understanding}. In this method, each active node independently chooses a slot out of a fixed number of slots uniformly at random, transmits in that slot with a fixed probability assigned to it and otherwise does not transmit. Phase 2 of the SRC$_S$ protocol uses the $BB$ method. In particular, phase 2 consists of a single trial of $\ell$ slots; each active node independently participates (respectively, does not participate) in the trial with probability $p$ (respectively, $1-p$). Also, each node that participates transmits in a slot selected uniformly at random from the $\ell$ slots (see Fig.~\ref{SRCs_window}).  The parameter $\ell$ is a function of the desired relative error $\epsilon$ and it is found from a numerical lookup table, which is constructed by executing the SRC$_S$ protocol for different values of $n$, and finding the value of $\ell$ required to achieve a given value of $\epsilon$~\cite{zhou2016understanding}.
Also,  the following parameter value  is used~\cite{zhou2016understanding}:
\begin{equation}
\label{EQ:SRCS:p}
p = \min{(1, 1.6 \ell/\tilde{n})}.
\end{equation} 
 Note that the expected fraction of empty slots, out of the $\ell$ slots, is ${(1- {p/\ell})} ^ {{n}}$. The protocol counts the number of empty slots, say $z$, out of the $\ell$ slots. The final estimate generated by the protocol is~\cite{zhou2016understanding}:
\begin{equation}
\label{EQ:SRCs:hatn}
\hat{n} = \frac{\ln({z/\ell})}{\ln{(1-{p/\ell})}}.
\end{equation} 

\subsection{Review of Node Cardinality Estimation Schemes for Heterogeneous M2M Networks Proposed in~\cite{kadam2017fast},~\cite{TechReport2018}}\label{GLOBECOM}
Two node cardinality estimation schemes are proposed in our prior work~\cite{kadam2017fast},~\cite{TechReport2018} by extending the LoF based protocol~\cite{qian2011cardinality} for obtaining separate estimates of the active node cardinalities of each node type in a heterogeneous M2M network with $T$ types of nodes. We now briefly review these two schemes since we use them as part of the estimation schemes proposed in this paper.

\begin{figure}
	\begin{center}
		\includegraphics[scale = 0.48]{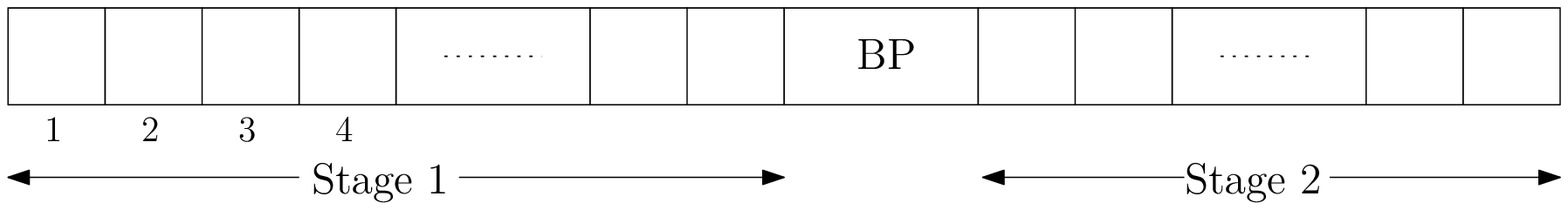}
		\caption{The figure shows the frame structure used in the 2-Stage Scheme (2-$SS$) proposed in~\cite{kadam2017fast},~\cite{TechReport2018}.}
		\label{Est_Window2}
	\end{center}
\end{figure}

The first scheme proposed in~\cite{kadam2017fast},~\cite{TechReport2018} consists of $3$ stages (see Fig.~\ref{Est_Window}) and the second scheme  consists of $2$ stages (except for $T=2$ and $T=3$) (see Fig.~\ref{Est_Window2}). So henceforth, we refer to them as ``The 3-Stage Scheme'' (3-$SS$) and ``The 2-Stage Scheme'' (2-$SS$) respectively. The active node cardinality estimate of each node type obtained using either of the schemes, 3-$SS$ and 2-$SS$, equals, and hence is as accurate as, the estimate that would have been obtained if the LoF based protocol were separately executed $T$ times to estimate the number of active nodes of each type. However, under mild conditions, the amounts of time needed by 3-$SS$ and 2-$SS$ to execute are much lower than the amount of time that would have been needed if the LoF based protocol were separately executed $T$ times. 

\subsubsection{The 3-Stage Scheme (3-$SS$)}\label{3SS}
\begin{figure}
	\begin{center}
		\includegraphics[scale = 0.43]{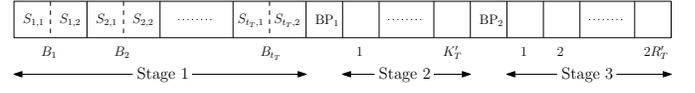}
		\caption{The figure shows the frame structure used in the 3-Stage Scheme (3-$SS$) proposed in~\cite{kadam2017fast},~\cite{TechReport2018} for the case $T=3$.}
		\label{Est_Window}
	\end{center}
\end{figure}

\begin{figure}
\centering
\resizebox{0.8\columnwidth}{!}{\includegraphics{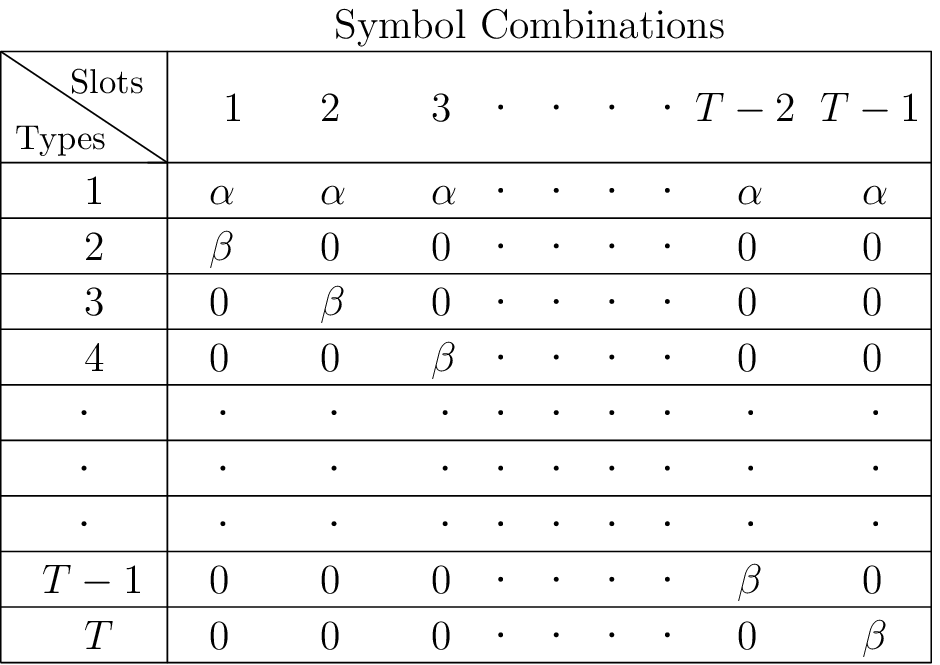}}
\caption{{ The figure shows the symbol combinations used by each type in the 3-Stage Scheme (3-$SS$) proposed in~\cite{kadam2017fast},~\cite{TechReport2018}. The symbol $0$ indicates ``no transmission''.}}
    \label{Sym_Combo3} 
\end{figure}

Let $n_{b, all}$ be the total number of nodes of $\mathscr{T}_b$ manufactured and $t_T = \ceil{\log_2 (\max (n_{1, all}, \ldots, n_{T, all}))}$.
Stage 1 of 3-$SS$ consists of $t_T$ blocks (see Fig.~\ref{Est_Window}). 
Each block, $B_h$, $h \in \{1,\ldots,t_T\}$, is divided into ($T-1$) slots $S_{h,1}$, \ldots, $S_{h,T-1}$. Each active node of each of the $T$ types independently chooses a block at random according to the distribution used in LoF based protocol (see \eqref{EQ:LoF:bar:pi}), i.e., the probability of choosing block $B_h$ is:
\begin{equation}
\label{EQ:pi}
p'_h = \left\{ \begin{array}{ll}
{1}/{2^{h}}, & \mbox{for } h = 1, \ldots, t_T - 1,\\
{1}/{2^{t_T - 1}}, & \mbox{for } h = t_T. \\
\end{array}
\right.
\end{equation} 
The symbol combinations used in this scheme are shown in Fig.~\ref{Sym_Combo3}. $\mathscr{T}_1$ active nodes whose chosen block is $B_h$ transmit symbol $\alpha$ in all ($T-1$) slots, i.e., $S_{h,1}$, \ldots, $S_{h,T-1}$, of block $B_h$. $\mathscr{T}_2$ (respectively, $\mathscr{T}_3, \ldots, \mathscr{T}_T$) active nodes whose chosen block is $B_h$ transmit symbol $\beta$ in slot $S_{h,1}$ (respectively, $S_{h,2}$, \ldots, $S_{h,T-1}$) and do not transmit in the other slots of block $B_h$. Stage 1 concludes with this. Now, it has been shown in~\cite{kadam2017fast},~\cite{TechReport2018} that if collisions occur in at most ($T-2$) slots of a given block $B_h$, then the set of types of nodes that transmitted in block $B_h$ can be unambiguously inferred by the BS. However, for some blocks of stage 1,  collisions in all ($T-1$) slots of the block $B_h$ may occur; in this case, the BS has  ambiguity about the types of nodes that transmitted in those particular blocks. To resolve the ambiguity, after the end of stage 1, the BS transmits a broadcast packet (BP), say BP$_1$ (see Fig.~\ref{Est_Window}), in which the list of the numbers of all blocks in which collisions in all ($T-1$) slots occurred is encoded. 

In stage 2, there are $K'_T$ slots, where $K'_T$ is the number of blocks in stage 1 in which collisions occurred in all ($T-1$) slots. For $i \in \{1, \ldots, K'_T$\}, in the $i^{th}$ slot of stage 2, $\mathscr{T}_1$ nodes that transmitted in the $i^{th}$ block of stage 1 in which collisions occurred in all ($T-1$) slots, transmit symbol $\alpha$. $\mathscr{T}_2, \ldots, \mathscr{T}_T$ nodes do not transmit in stage 2. Now, it is easy to see that at the end of stage 2, the BS unambiguously knows the set of block numbers of stage 1 in which $\mathscr{T}_1$ nodes transmitted. However,  if in stage 2, there are collisions in some of the slots, ambiguity remains with the BS on whether $\mathscr{T}_2, \ldots, \mathscr{T}_T$ nodes transmitted in the corresponding blocks of stage 1. To resolve this ambiguity, after the end of stage 2, the BS transmits a BP, say BP$_2$ (see Fig.~\ref{Est_Window}), in which is encoded, the list of block numbers of stage 1 for which collisions occurred in the corresponding slots of stage 2. Suppose there are $R'_T$ blocks in this list. 

In stage 3, $(T-1)R'_T$ slots are used. For $i \in \{1,\ldots,R'_T\}$,  $\mathscr{T}_2$ (respectively, $\mathscr{T}_3, \ldots, \mathscr{T}_T$) active nodes corresponding to the $i^{th}$ block in the above list transmit symbol $\beta$ in the $((i-1)(T-1)+1)^{th}$ (respectively, $((i-1)(T-1)+2)^{th}$, \ldots, $(i(T-1))^{th}$) slot of stage 3. It is easy to see that for each $b \in \{1, \ldots, T\}$, at the end of stage 3, the BS unambiguously knows the set of block numbers of stage 1 in which $\mathscr{T}_b$  nodes transmitted. 

For $b \in \{1, \ldots, T\}$, let $j_b$ be the smallest number $j$ such that no $\mathscr{T}_b$ node transmitted in the $j^{th}$ block of stage $1$.\footnote{If at least one $\mathscr{T}_b$ node transmitted in all the $t_T$ blocks of stage 1, then $j_b = t_T$.} Then the estimate of the number of active nodes of $\mathscr{T}_b$ is $1.2897 \times 2^{j_b-1}$ (see \eqref{EQ:LoF:estimate:nhat}). 

More generally, suppose the above 3-$SS$ scheme is independently executed $M$ times. 
For $m \in \{1, \ldots, M\}$ and $b \in \{1, \ldots, T\}$, let $j_b(m)$ be the smallest number $j$ such that no $\mathscr{T}_b$ node transmitted in the $j^{th}$ block of stage $1$ in the $m^{th}$ trial.\footnote{If at least one $\mathscr{T}_b$ node transmitted in all the $t_T$ blocks of stage 1 in the $m^{th}$ trial, then $j_b(m) = t_T$.} Then the estimate of the number of active nodes of $\mathscr{T}_b$ is $1.2897 \times 2^{\Sigma_{m=1}^M (j_b(m) - 1)/M}$ (see \eqref{EQ:LoF:estimate:nhat}). 

\begin{figure}
\centering
\resizebox{0.9\columnwidth}{!}{\includegraphics{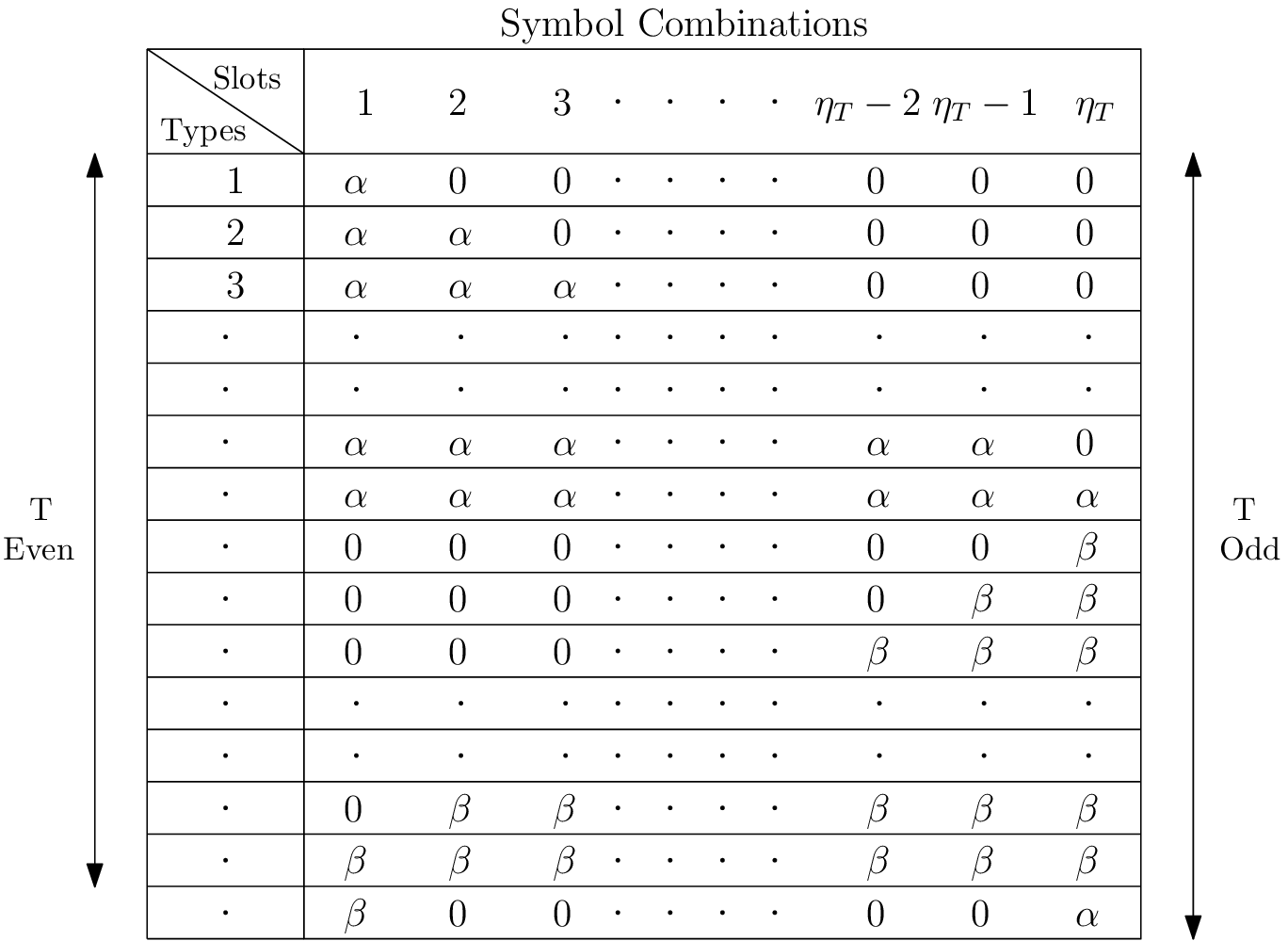}}
\caption{{ The figure shows the symbol combinations used by each type in the 2-Stage Scheme (2-$SS$) proposed in~\cite{kadam2017fast},~\cite{TechReport2018}. $\eta_T = T/2$ if $T$ is even and $\eta_T = (T-1)/2$ if $T$ is odd. The symbol $0$ indicates ``no transmission''.}}
    \label{Sym_Combo} 
\end{figure}

\subsubsection{The 2-Stage Scheme (2-$SS$)}\label{2SS}
For $T=2$ and $T=3$, 2-$SS$ is identical to 3-$SS$. We now explain the operation of 2-$SS$ for $T \ge 4$. For $T \ge 4$,  2-$SS$ is a more sophisticated scheme than 3-$SS$ and has only two stages. Stage 1 of 2-$SS$ consists of $t_T$ blocks (see Fig.~\ref{Est_Window2}). 
Each block, $B_h$, $h \in \{1,\ldots,t_T\}$, is divided into ($T/2$) slots if $T$ is even and ($T-1$)$/2$ slots if $T$ is odd. Each active node of each of the $T$ types independently chooses a block number, $h$, at random using the distribution given in \eqref{EQ:pi}. The symbol combinations used in 2-$SS$ are shown in Fig.~\ref{Sym_Combo}.\footnote{In particular, $\mathscr{T}_1$ active nodes whose chosen block is $B_h$ transmit symbol $\alpha$ in the first slot of $B_h$ and do not transmit in the other slots of $B_h$. $\mathscr{T}_2$ active nodes whose chosen block is $B_h$ transmit symbol $\alpha$ in the first two slots of $B_h$ and do not transmit in the other slots of $B_h$. If $T$ is even (respectively, odd), $\mathscr{T}_{T/2}$ (respectively, $\mathscr{T}_{(T-1)/2}$) active nodes whose chosen block is $B_h$ transmit symbol $\alpha$ in all the slots of $B_h$. If $T$ is even (respectively, odd), $\mathscr{T}_{(T/2)+1}$ (respectively, $\mathscr{T}_{((T-1)/2)+1}$) active nodes whose chosen block is $B_h$ transmit symbol $\beta$ in the last slot of $B_h$ and do not transmit in the other slots of $B_h$ and so on. Finally, if $T$ is even, then $\mathscr{T}_T$ active nodes whose chosen block is $B_h$ transmit symbol $\beta$ in all the slots of $B_h$ and if $T$ is odd, then $\mathscr{T}_T$ active nodes transmit symbol $\beta$ in the first slot and symbol $\alpha$ in the last slot of $B_h$ and do not transmit in the other slots of $B_h$.} Now, it is easy to see that if collisions do not occur in any of the slots of a block $B_h$, then the set of types of nodes that transmitted in block $B_h$ can be unambiguously inferred by the BS. 
In case of collisions in at least one slot, but not all slots, of a block $B_h$, ambiguity may remain and it is resolved in stage 2.\footnote{For example, consider $T=4$. In stage 1, if slot 1 results in $\beta$ and slot 2 results in $C$, then the BS unambiguously infers that at least one node of $\mathscr{T}_3$ and exactly one node of $\mathscr{T}_4$ are active, and all $\mathscr{T}_1$ and $\mathscr{T}_2$  nodes are inactive. Stage 2 is not required in this case. Similarly, if slot 1 results in $\alpha$ and slot 2 results in $C$, then the BS infers that at least one node of $\mathscr{T}_3$ is active, no node of $\mathscr{T}_4$  is active, and exactly  one node of either $\mathscr{T}_1$  or $\mathscr{T}_2$  is active. Stage 2 is required in this case to resolve the ambiguity about whether a $\mathscr{T}_1$  or $\mathscr{T}_2$ node is active.} In case of collisions in all the slots of a block $B_h$, ambiguity remains about the activity or inactivity of each of the node types. In this case, the set of all node types $\{1,\ldots,T\}$ is divided into smaller groups and each of these groups  recursively uses the stage 1 protocol in stage 2 to resolve the ambiguity.\footnote{For example, consider $T=5$. If the block result  $C C$ occurs in stage 1, then the set of node types $\{1, \ldots, 5\}$ is divided into two groups: $\{1, 2, 3\}$ and $\{4, 5\}$. For the first (respectively, second) group, the stage 1 scheme for $T = 3$ (respectively, $T=2$) node types is (recursively) used in stage 2 to resolve the ambiguity.} 

A broadcast packet (BP) is sent by the BS after stage 1, which contains instructions that the active nodes should follow to resolve the remaining ambiguity, if any, in stage 2. 

It has been shown in~\cite{kadam2017fast},~\cite{TechReport2018} that for each $b \in \{1, \ldots, T\}$, at the end of stage 2, the BS unambiguously knows the set of block numbers of stage 1 in which $\mathscr{T}_b$  nodes transmitted.

Estimates of the number of active nodes of $\mathscr{T}_b$, $b \in \{1, \ldots, T\}$, are computed under the above 2-$SS$ scheme similar to their computation under the 3-$SS$ scheme-- see the last two paragraphs of Section~\ref{3SS}.

\section{Proposed Node Cardinality Estimation Schemes for Heterogeneous M2M Networks}\label{EstScheme}
We now describe the proposed schemes, which are extensions of the SRC$_S$ protocol for estimating the number of active nodes of each type in the model with a BS and $T$ different types of nodes in its range described in Section~\ref{nwmodel}. The proposed schemes are the Heterogeneous SRC$_S$-1 scheme (HSRC-1) and  the Heterogeneous SRC$_S$-2 scheme (HSRC-2) and both consist of two phases-- they correspond to the two phases of the SRC$_S$ protocol (see Section~\ref{SubSec_Est}).

Recall from Section~\ref{SubSec_Est} that phase 1 of the SRC$_S$ protocol is a series of $M^{\prime}$ independent trials of the LoF based protocol. While extending the SRC$_S$ protocol for node cardinality estimation in a heterogeneous network with $T$ types of nodes, one possibility is to separately execute phase 1 of the SRC$_S$ protocol $T$ times for estimating the active node cardinalities of the $T$ node types.\footnote{Note that this would require execution of $M^{\prime}$ independent trials of the LoF based protocol for each node type, i.e., a total of $M^{\prime} T$ independent trials, in phase 1.}  However, since it is shown in~\cite{kadam2017fast},~\cite{TechReport2018} that, under mild conditions, a trial of  3-$SS$  (respectively, 2-$SS$) takes less time compared to $T$ separate executions of a trial of the LoF based protocol for estimating the active node cardinalities of the $T$ types of nodes, we use a series of $M^{\prime}$ independent executions of 3-$SS$ (respectively, 2-$SS$) in phase 1 of HSRC-1 (respectively, HSRC-2). At the end of phase 1 of HSRC-1 or HSRC-2,  we obtain rough estimates, say $\tilde{n}_1,\ldots, \tilde{n}_T$, of the numbers of active nodes of $\mathscr{T}_1,\ldots, \mathscr{T}_T$ respectively. Note that these estimates are the same as those that would have been obtained if phase 1 of the SRC$_S$ protocol were separately executed $T$ times for obtaining rough estimates of the active node cardinalities of the $T$ node types.   

Next, recall from Section~\ref{SubSec_Est} that phase 2 of the SRC$_S$ protocol consists of a single  $BB$ (balls-and-bins) trial. The number of slots, $\ell$, in the trial depends on the desired relative error $\epsilon$ (see Section~\ref{nwmodel}). Since the value of $\epsilon$ is the same for all the $T$ node types (see Section~\ref{nwmodel}), the  length, $\ell$, of the trial is the same for all the $T$ node types. For $b \in \{1, \ldots, T\}$, let (see \eqref{EQ:SRCS:p}):
\begin{equation}
\label{EQ:pb}
p_b = \min \left({1,\frac{1.6\ell}{\tilde{n}_b}} \right).
\end{equation} 
Now, one possible approach to execute phase 2 of the proposed schemes is to separately execute $T$ $BB$ trials-- one trial for each of the $T$ node types;  note that in the trial for  $\mathscr{T}_b$ nodes,  the probability $p_b$ in \eqref{EQ:pb} is used as the probability with which each active node participates. This approach requires a total of $T\ell$ time slots to execute. We refer to this approach as ``$T$-$Rep$-$BB$''.  

An alternative approach to execute phase 2 of the proposed schemes is to use  the method ``3-$SS$-$BB$'' or the method ``2-$SS$-$BB$'', which are as follows. 
 3-$SS$-$BB$ (respectively, 2-$SS$-$BB$) is the method of executing 3-$SS$ (respectively, 2-$SS$) similar to the scheme described in Section~\ref{3SS} (respectively, Section~\ref{2SS}), with the change that in stage 1, $\ell$ blocks are used (instead of $t_T$ blocks) and for $b \in \{1, \ldots, T\}$, each node of $\mathscr{T}_b$ independently transmits with the probability $p_b$ in \eqref{EQ:pb} (instead of $p'_h, h \in \{1, \ldots, t_T\}$, see \eqref{EQ:pi}) in a block chosen uniformly at random from the $\ell$ blocks and does not transmit with probability $1 - p_b$. Ambiguities about the sets of types of nodes that transmitted in different blocks of stage 1 are resolved in stages 2 and 3 (respectively, in stage 2) of 3-$SS$-$BB$ (respectively, 2-$SS$-$BB$) as explained in Section~\ref{3SS} (respectively, Section~\ref{2SS}). Hence, in case of 3-$SS$-$BB$ (respectively, 2-$SS$-$BB$), at the end of stage 3 (respectively, stage 2), the BS unambiguously knows the sets, say $\mathcal{I}_1, \ldots, \mathcal{I}_T$, of block numbers of stage 1 in which $\mathscr{T}_1, \ldots, \mathscr{T}_T$ nodes respectively transmitted. From the sets $\mathcal{I}_1,  \ldots, \mathcal{I}_T$, for each $b \in \{1, \ldots, T\}$, $z_b$, which is the number of slots that would have been empty if phase $2$ of the SRC$_S$ protocol were executed for $\mathscr{T}_b$ nodes, can be deduced. For each $b \in \{1, \ldots, T\}$, the final estimate of the number of active nodes of $\mathscr{T}_b$ is calculated at the end of phase 2 as $\hat{n}_b=\ln(z_b/\ell)/\ln(1-p_b/\ell)$ (see \eqref{EQ:SRCs:hatn}). 

Note that irrespective of which of the above approaches--viz., $T$-$Rep$-$BB$,  3-$SS$-$BB$ or 2-$SS$-$BB$-- is used, \emph{the final node cardinality estimate, $\hat{n}_b$, of each type $b \in \{1, \ldots, T\}$, obtained using the proposed schemes equals, and hence is as accurate as, the estimate that would have been obtained if the SRC$_S$  protocol were separately executed $T$ times to estimate the number of active nodes of each type.}      

In HSRC-1 (respectively, HSRC-2), a series of $M^{\prime}$ independent executions of 3-$SS$ described in Section~\ref{3SS} (respectively, 2-$SS$ described in Section~\ref{2SS}) is used in phase 1 and depending on a certain condition, either 3-$SS$-$BB$ or $T$-$Rep$-$BB$ (respectively, either 2-$SS$-$BB$ or $T$-$Rep$-$BB$) is used in phase 2. This condition for HSRC-1 is derived analytically in Section~\ref{Boundary}.

\section{Phase 2 of  HSRC-1}
\label{SC:HSRC1:phase2:condition}
In order to minimize the execution time of phase 2 of HSRC-1, we have derived a condition, which, if satisfied, we use 3-$SS$-$BB$, else we use $T$-$Rep$-$BB$ in phase 2 of HSRC-1. In Section~\ref{p2_3stage}, we compute the expected numbers of time slots required if  $T$-$Rep$-$BB$ is used and if 3-$SS$-$BB$ is used in phase 2 of HSRC-1  and we use these results in Section~\ref{Boundary} to derive the condition using which we decide as to which approach to use in phase 2 of HSRC-1. 
\subsection{Expected Number of Slots Required in Phase 2 of HSRC-1}\label{p2_3stage}
Recall from Section~\ref{EstScheme} that if $T$-$Rep$-$BB$ is used, then   $T\ell$  slots are required in phase 2 of HSRC-1. Now we compute the expected number of slots required in phase 2 assuming that 3-$SS$-$BB$ is used. 

The number of slots required in stage 1 is $(T-1)\ell$ (see Section~\ref{EstScheme}). Let $K_T$ (respectively, $(T-1)R_T$) be the number of slots required in stage 2 (respectively, stage 3). Let ${S}_{h,1}^r$ (respectively, ${S}_{h,2}^r, \ldots, {S}_{h,T-1}^r$), $h \in \{1, \ldots, \ell\}$, represent the result (collision, success or empty slot) of the first (respectively, second, \ldots, $(T-1)^{th}$) slot of block $B_h$ of stage 1. Also, let {I}$_{\upsilon}$ denote the indicator random variable corresponding to event $\upsilon$, i.e., {I}$_{\upsilon}$ is 1 if $\upsilon$ occurs, else it is 0.
 
From Sections~\ref{3SS} and~\ref{EstScheme}, it is easy to see that  $K_T$ = $\Sigma^{\ell}_{{h} = 1} {I}_{\{{S}_{h,1}^r = C, \ldots, {S}_{h,T-1}^r = C\} }$, where $C$ denotes collision. So:
\begin{equation}
E(K_T) = \sum^{\ell}_{{h} = 1} {P}({S}_{h,1}^r = C, \ldots, {S}_{h,(T-1)}^r = C).
\label{eq1}
\end{equation}

The conditions under which collisions occur in all $(T-1)$ slots of block $B_h$, $h \in \{1, \ldots, \ell\}$, are as follows: 
\begin{enumerate}
\item At least two nodes of $\mathscr{T}_1$ transmit in block $B_h$.
\item Exactly one node of $\mathscr{T}_1$ and at least one node each of $\mathscr{T}_2, \ldots, \mathscr{T}_T$ transmit in block $B_h$.
\item At least two nodes each of $\mathscr{T}_2, \ldots, \mathscr{T}_T$ and none of $\mathscr{T}_1$ transmit in block $B_h$.
\end{enumerate}

Let $Q_1(h)$, $Q_2(h)$, and $Q_3(h)$ denote the probabilities of the events in $1)$, $2)$, and $3)$ respectively. Since the probability of selecting  a block $B_h$ by the nodes of a given $\mathscr{T}_b$ is the same for all the blocks $B_h$ irrespective of $h$, we can write: $Q_{j}(h) = Q_{j}$, $j \in\{1,2, 3\}, \ h \in \{1, \ldots, \ell\}$.  
Hence:
\begin{equation}
{P}({S}_{h,1}^r = C, \ldots, {S}_{h,(T-1)}^r = C) = Q_{1}+Q_{2}+Q_{3}.
\label{equ2}
\end{equation}
Also:
\begin{align} 
Q_1 & = 1 - u_1{(n_1)} - v_1{(n_1)},
\label{eqQ1}\\
Q_2  & = v_1{(n_1)} \prod_{b=2}^{T}\left(1 -  u_b(n_b)\right),
\label{eqQ2}\\
Q_3  & = u_1{(n_1)} \prod_{b=2}^{T}\left(1 - u_b(n_b) - v_b(n_b)\right),
\label{eqQ3}
\end{align}
where $u_b(n_b), b \in \{1, \ldots, T\}$, is the probability that none of the nodes out of the $n_b$ nodes of $\mathscr{T}_b$ select a given block and $v_b{(n_b)}$ is the probability that exactly one node out of the $n_b$ nodes of $\mathscr{T}_b$ selects a given block. 
So: 
\begin{align} 
u_b{({n}_b)} & =  \left(1 - \frac{p_b}\ell \right)^{{n}_b},
\label{equnb}\\
v_b({n}_b) & = {{n}_b} \frac{p_b}\ell \left( 1 - \frac{p_b}\ell \right)^{{{n}_b} -1},
\label{eqvnb}
\end{align} 
where $p_b = \min{\left(1, \frac{1.6 \ell}{{\tilde{n}}_b}\right)}$ (see~\eqref{EQ:pb}). 
By~\eqref{eq1} and~\eqref{equ2}:
\begin{equation}
\label{EQ:EK}
E(K_T) = \ell (Q_1 + Q_2+ Q_3). 
\end{equation}
Also: 
\begin{equation}
\label{EQ:ER}
E(R_T) = \ell Q_1,  
\end{equation}
since in stage 3, only those nodes of $\mathscr{T}_2, \ldots, \mathscr{T}_T$ transmit for which collisions occurred  in all the slots of the corresponding blocks of stage 1 due to two or more $\mathscr{T}_1$ nodes transmitting (see Sections~\ref{3SS} and~\ref{EstScheme}). The expected total number of slots required in phase 2 of HSRC-1, when 3-$SS$-$BB$ is used in phase 2, is $(T-1) \ell + E(Z_{BP})+ E(K_T) + (T-1) E(R_T)$, where $Z_{BP}$ is the number of slots required by the broadcast packets BP$_1$ and BP$_2$ (see Fig.~\ref{Est_Window}).

\subsection{Condition Used to Select Approach to be Used in Phase 2 of HSRC-1}\label{Boundary}
From the description of 3-$SS$-$BB$ in Sections~\ref{3SS} and~\ref{EstScheme}, it can be seen that in stage 1, if a $\mathscr{T}_1$ node chooses a block $B_h, h \in \{1, \ldots, \ell\}$, it transmits in all the slots ${S_{h,1}}, \ldots, {S_{h,T-1}}$, whereas if a node of $\mathscr{T}_b,  b \in \{2, \ldots, T\}$, selects block $B_h$, it transmits only in one slot, viz., ${S_{h,b-1}}$. So, the number of collisions due to $\mathscr{T}_1$ nodes is high compared to those due to $\mathscr{T}_b$ nodes, $b \geq 2$. Also, clearly the numbers of slots required in stage 2 and stage 3 increase with the number of collisions in stage 1. Therefore, the numbers of slots required in stage 2 and stage 3 increase rapidly whenever the number of $\mathscr{T}_1$ nodes is increased. Hence, we develop a condition on $\tilde{n}_1$ (see Section~\ref{EstScheme}): if it is less than a certain value, we use 3-$SS$-$BB$, else we use $T$-$Rep$-$BB$ in phase 2 of HSRC-1. It is possible to check whether the condition holds because we already have a rough estimate of ${n}_{b}, b \in \{1, \ldots, T\}$, i.e., $\tilde{n}_{b}$ (see \eqref{EQ:LoF:estimate:nhat}), from phase 1 using which it can be checked whether the condition holds. Hence, \emph{we use $\tilde{n}_{b}$ instead of $n_b$ throughout this section}.

To derive the condition, note that the use of 3-$SS$-$BB$ is profitable only if the number of slots required when it is used is not more than $T\ell$ (which is the number of slots required by $T$-$Rep$-$BB$); also, note that the number of slots required increases with  increase in $\tilde{n}_{2}, \ldots, \tilde{n}_{T}$. So, we keep $\tilde{n}_{2}, \ldots, \tilde{n}_{T}$ very large, i.e., we let them approach infinity, and  we derive a condition on $\tilde{n}_{1}$ for which the expected number of slots required when 3-$SS$-$BB$ is used is not more than $T\ell$. This ensures that when this condition is satisfied, the expected number of slots required by 3-$SS$-$BB$ is $\leq T\ell$ regardless of the values of $\tilde{n}_{2}, \ldots, \tilde{n}_{T}$. 
Now, recall from Section~\ref{p2_3stage} that the expected number of slots required by 3-$SS$-$BB$ is $(T-1) \ell + E(Z_{BP})+ E(K_T) + (T-1) E(R_T)$. So the required condition is: $(T-1) \ell + E(Z_{BP}) + E(K_T) + (T-1) E(R_T) \leq T \ell$, i.e., 
\begin{equation}
\label{EQ:cond1}
E(K_T) +  E(Z_{BP}) + (T-1)E(R_T) \leq \ell. 
\end{equation}
Note that $\ell/S_W  +  K_T/S_W \le Z_{BP} = \ceil{\ell/S_W} + \ceil{K_T/S_W}$\footnote{The first term, $\ceil{\ell/S_W}$,  equals the length of the BP$_1$ sent by the BS after stage 1 (see Fig.~\ref{Est_Window}) in terms of number of slots. This BP contains a string of $\ell$ bits  that indicates the results of all $\ell$ blocks of stage 1. In particular, if the bit in the $i^{th}$ position in the bit string is $1$ (respectively, $0$), then this  indicates that stage 2 is required (respectively, not required) to resolve the ambiguity regarding the node types, if any, that transmitted in block $B_i$.  Similarly, the second term, $\ceil{K_T/S_W}$, equals the length of the BP$_2$ sent by the BS after stage 2 (see Fig.~\ref{Est_Window}) in terms of number of slots. Note that this BP contains a bit string that indicates the results of the $K_T$ slots of stage 2.} $< \ell/S_W  + 1 +  K_T/S_W + 1$,\footnote{This inequality follows from the fact that $x \le \ceil{x} < x+1, \forall x \in \mathscr{R}$.} where $S_W$ denotes the slot width in bits. So $\ell/S_W  +  E(K_T)/S_W \le E(Z_{BP}) < E(K_T)/S_W  + \ell/S_W + 2$.
Hence, a sufficient (respectively, necessary) condition for \eqref{EQ:cond1} to hold is \eqref{EQ:cond2} (respectively, \eqref{EQ:cond3}):
\begin{align}
\label{EQ:cond2}
(1 + 1/S_W) E(K_T)  + (T-1)E(R_T) & \leq \ell (1 - 1/S_W) - 2, \\
\label{EQ:cond3}
(1 + 1/S_W) E(K_T)  + (T-1)E(R_T) & \leq \ell (1 - 1/S_W). 
\end{align}

Since  $\tilde{n}_2,\ldots, \tilde{n}_T$ are assumed to be very large, they are $>> 1.6 \ell$.
Therefore $p_2 = \min \left({1,\frac{1.6\ell}{\tilde{n}_2}} \right)=\left(\frac{1.6\ell}{\tilde{n}_{2}}\right)$ and similarly $p_3$ = $\left(\frac{1.6\ell}{\tilde{n}_{3}}\right),\ldots, p_T = \left(\frac{1.6\ell}{\tilde{n}_{T}}\right)$.  
By using~\eqref{equnb} and~\eqref{eqvnb}, for very large values of $\tilde{n}_2$, we get: 
\begin{align}
\lim_{\tilde{n}_2\to\infty} u_2(\tilde{n}_2)  & = \lim_{\tilde{n}_2\to\infty} \Big(1-\frac{1.6}{{\tilde{n}}_{2}}\Big)^{\tilde{n}_{2}} = {e}^{-1.6},\\
\lim_{\tilde{n}_2\to\infty} v_2(\tilde{n}_2)  & = \lim_{{\tilde{n}}_{2}\to\infty} 1.6\Big(1-\frac{1.6}{{\tilde{n}}_{2}}\Big)^{\tilde{n}_{2}} = 1.6{e}^{-1.6}.
\end{align}
Similarly for very large values of $\tilde{n}_3,\ldots,\tilde{n}_T$, we get:  $\lim_{\tilde{n}_3\to\infty}  u_3(\tilde{n}_3) = \ldots = \lim_{\tilde{n}_T\to\infty}  u_T(\tilde{n}_T) ={e}^{-1.6}$ and $\lim_{\tilde{n}_3\to\infty}  v_3(\tilde{n}_3) = \ldots = \lim_{\tilde{n}_T\to\infty}  v_T(\tilde{n}_T) = 1.6{e}^{-1.6}$.

First, we introduce some notation. For simplicity, let us assume $S_W = 6$~\cite{kodialam2007anonymous}, which is a typical value in practice and let $G_1(T) = (1 + 6T) - 7(0.4751)^{T-1}$ and $G_2(T) = (1 + 6T) - 7(0.7981)^{T-1}$. Let $f(x, T) = (0.366)^{x}\left(G_1(T)+  xG_2(T)\right)$ and $f_1(x,T) = (0.3679)^{x}\left(G_1(T) + xG_2(T)/0.99\right)$ for $x > 0$. Assuming that $T \le 50$, which would typically be the case in practice, $f(x, T)$ and $f_1(x, T)$ are decreasing functions of $x$ for $x > 0$.\footnote{It can be easily shown that: $\frac{\partial}{\partial x} f(x, T) = -(0.366)^x (1.005 G_1(T) - G_2(T) + 1.005xG_2(T)) < 0$, $\forall x > 0$ since $G_1(T) > G_2(T) > 0, \forall T$. Similarly, it is easy to show that: $\frac{\partial}{\partial x} f_1(x, T) = -(0.3679)^x \left(G_1(T) - \frac{G_2(T)}{0.99} + x\frac{G_2(T)}{0.99}\right) < 0$, $\forall x > 0$ since $G_1(T) > \frac{G_2(T)}{0.99} > 0, \forall T \le 50$.} Let $\zeta_1(T)$ (respectively, $\zeta_2(T)$) be the largest  (respectively, smallest) value of $x$ such that $f(x, T) \ge 6T - 3.88$ (respectively, $f_1(x, T) < 6T -4$), $\forall x \leq \zeta_1(T)$ (respectively, $\forall x \geq \zeta_2(T)$). 

\begin{remark}\label{zeta_remark}
Note that in practice, the values of $\zeta_1(T)$ and $\zeta_2(T)$ can be readily computed as follows. For a fixed $T$, $f(x, T)$ (respectively, $f_1(x, T)$) can be plotted with respect to $x >0$; the value of $x$ where the function equals $6T - 3.88$ (respectively, $6T - 4$) can be taken as $\zeta_1(T)$ (respectively, $\zeta_2(T)$). 
\end{remark}

Now, we consider the cases (I) $\tilde{n}_{1} < 1.6\ell$ and (II) $\tilde{n}_{1} \geq 1.6\ell$ separately, and in each case, we investigate as to which values of  $\tilde{n}_{1}$ satisfy the condition in \eqref{EQ:cond1}.  
The proofs of  the following propositions (Proposition~\ref{Case1_Cond}--\ref{PN:Case2}) are relegated to the Appendix.
 
\subsubsection{Case I: $\tilde{n}_{1} < 1.6\ell$}\label{Case1}
This implies $p_1 = \min \left({1,\frac{1.6\ell}{\tilde{n}_1}} \right) = 1$.
\begin{proposition}
\label{Case1_Cond}
When $\tilde{n}_{1} < 1.6\ell$, a sufficient (respectively, necessary) condition  for \eqref{EQ:cond1} to hold is \eqref{EQ:NewCond6} (respectively, \eqref{EQ:NewCond6a}):
\begin{align}
\label{EQ:NewCond6}
 G_1(T) \Big(1 - \frac{1}\ell\Big)^{{\tilde{n}}_{1}}  +   G_2(T) \frac{{\tilde{n}}_{1}}\ell\Big(1 - \frac{1}\ell\Big)^{{\tilde{n}}_{1}-1}  & \ge 6T - 4 + 12/\ell. \\
 \label{EQ:NewCond6a}
 G_1(T) \Big(1 - \frac{1}\ell\Big)^{{\tilde{n}}_{1}}  +   G_2(T) \frac{{\tilde{n}}_{1}}\ell\Big(1 - \frac{1}\ell\Big)^{{\tilde{n}}_{1}-1}  & \ge 6T - 4.
\end{align}
\end{proposition}

\begin{proposition}
\label{PN:Case1}
Assume that $\ell \geq 100$ and $T \le 50$. Inequality~\eqref{EQ:NewCond6} holds when ${\tilde{n}}_{1} \leq  \zeta_1(T) \ell$. Also, inequality~\eqref{EQ:NewCond6a} does not hold when $\zeta_2(T) \ell \leq {\tilde{n}}_{1} < 1.6\ell$.
\end{proposition}

Assuming that $l \geq 100$ and $T \le 50$ (which would most likely be the case in practice), Proposition~\ref{PN:Case1} shows that whenever ${\tilde{n}}_{1} \leq \zeta_1(T) \ell$ (respectively, $\zeta_2(T) \ell \leq {\tilde{n}}_{1} < 1.6\ell$), \eqref{EQ:cond1} holds (respectively, does not hold) and hence 3-$SS$-$BB$ takes less  (respectively, more) time on average than $T$-$Rep$-$BB$ in phase 2 of HSRC-1.

\subsubsection{Case II: $\tilde{n}_{1} \ge 1.6\ell$}
This implies $p_1 = \min \left({1,\frac{1.6\ell}{\tilde{n}_1}} \right) = 1.6\ell/\tilde{n}_{1}$.
\begin{proposition}
\label{Case2_Cond}
When $\tilde{n}_{1} \ge 1.6\ell$, a necessary condition for \eqref{EQ:cond1} to hold  is:
\begin{align}
\label{EQ:NewCond8}
G_1(T) \left(1 - \frac{1.6}{\tilde{n}_{1}}\right)^{{\tilde{n}}_{1}}   +  G_2(T) 1.6\left(1 - \frac{1.6}{\tilde{n}_{1}}\right)^{{\tilde{n}}_{1}-1}   \ge 6T - 4.
\end{align}
\end{proposition}

\begin{proposition}
\label{PN:Case2}
Inequality \eqref{EQ:NewCond8} does not hold when $\tilde{n}_{1} \ge 1.6\ell$ and $\ell \geq 100$.
\end{proposition}

Proposition~\ref{PN:Case2} shows that when $\ell \geq 100$, the condition in \eqref{EQ:NewCond8}, and hence that in \eqref{EQ:cond1}, does not hold for any value of ${\tilde{n}}_{1} \ge 1.6\ell$. Thus, $T$-$Rep$-$BB$ takes less time on average than 3-$SS$-$BB$ in phase 2 of HSRC-1 for all values of ${\tilde{n}}_{1} \ge 1.6\ell$.

In summary, the analysis of cases I and II shows that when $\tilde{n}_1 \leq \zeta_1(T)  \ell$ (respectively, $\tilde{n}_1 \geq \zeta_2(T) \ell$), 3-$SS$-$BB$ takes less (respectively, more) time on average than $T$-$Rep$-$BB$ in phase 2 of HSRC-1. It is unclear from the analysis as to which technique takes less time when $\tilde{n}_1 \in (\zeta_1(T)  \ell, \zeta_2(T) \ell)$. This question is addressed via simulations in Section~\ref{Simu}.

\section{Performance Analysis}\label{Analysis}  
In this section, the expected number of time slots required by HSRC-1  to execute and the expected energy consumption of a node under  the  scheme in various cases are mathematically analysed.

\subsection{Expected Number of Slots Required by HSRC-1}\label{Analysis_HSRC1} 
Recall from Section~\ref{EstScheme} that in phase 1 of HSRC-1, we use 3-$SS$ and in phase 2, based on the condition obtained in Section~\ref{Boundary}, we use either $T$-$Rep$-$BB$ or 3-$SS$-$BB$.
When $T$-$Rep$-$BB$ is used in phase 2, it takes $T \ell$ slots to execute in that phase. 
Now we compute the expected number of slots required by 3-$SS$ (say $\Lambda_{I}$) and 3-$SS$-$BB$ (say $\Lambda_{II}$) to execute. Recall from Section~\ref{3SS} that 3-$SS$ consists of three stages and two BPs, BP$_1$ and BP$_2$. Also, stage 1 (respectively, stage 2, stage 3) takes $(T-1)t_T$ (respectively,  $E[K'_T]$,  $(T-1)E[R'_T]$) slots to execute, and BP$_1$ (respectively, BP$_2$)  takes $\ceil{t_T/S_W}$ (respectively, $\ceil{E[K'_T]/S_W}$) slots to execute. Thus, the expected number of slots required by 3-$SS$ to execute is:
\begin{multline}
\label{eq_P1_3SS}
\Lambda_I =   (T-1)t_T  + \ceil{t_T/S_W} \\ + E[K'_T] +  \ceil{E[K'_T]/S_W}   + (T-1)E[R'_T].
\end{multline}
Closed form expressions for $E[K'_T]$ and $E[R'_T]$ can be found in our prior work~\cite{TechReport2018}.

Next, recall from Section~\ref{p2_3stage} that  3-$SS$-$BB$ also has three stages and two BPs, BP$_1$ and BP$_2$. Stage 1 (respectively, stage 2, stage 3) takes $(T-1)\ell$ (respectively,  $E[K_T]$,  $(T-1)E[R_T]$) slots to execute. Also,  BP$_1$ (respectively, BP$_2$) takes $\ceil{\ell/S_W}$ (respectively, $\ceil{E[K_T]/S_W}$) slots to execute. Hence, the expected number of slots required by 3-$SS$-$BB$ to execute is: 
\begin{multline}
\label{eq_P1_3BB}
\Lambda_{II} = (T-1)\ell  + \ceil{\ell/S_W} \\ + E[K_T]  +  \ceil{E[K_T]/S_W} + (T-1)E[R_T].
\end{multline}
Closed form expressions for $E[K_T]$ and $E[R_T]$ are provided by \eqref{EQ:EK} and \eqref{EQ:ER} respectively.

\subsection{Expected Energy Consumption of a Node under  HSRC-1}
In this subsection, first we compute the expected energy consumption of a  node under 3-$SS$, 3-$SS$-$BB$, and $T$-$Rep$-$BB$. Then we find the expected energy consumption under HSRC-1.  Let $\gamma_\tau$, $\gamma_\rho$, and $\gamma_\iota$ be the energy spent by a node per slot in the transmission state, reception state, and idle state respectively.  (We assume that the energies required to transmit the symbols $\alpha$ and $\beta$ are the same.) If a node is inactive in a frame, then its energy consumption is $\gamma_\iota$ per slot throughout the frame. So in the rest of this section, we find the energy consumption of active nodes in a given frame.

\subsubsection{Expected Energy Consumption of a Node in 3-$SS$}
\label{SSSC:3SS:energy:consumption}
For each $b \in \{1, \ldots, T\}$, let $\mathscr{N}_{b}$ be the set of active nodes of $\mathscr{T}_b$ and  $w_b$ be any node from $\mathscr{N}_{b}$ that selects block $h$ in stage 1. Also, for a given $h \in \{1, \ldots, t_T\}$, let $\mathscr{E}_{b, \tau}^{(h)}$, $\mathscr{E}_{b, \rho}^{(h)}$, and $\mathscr{E}_{b, \iota}^{(h)}$ be the total energy consumed by a node of $\mathscr{T}_{b}, b \in \{1, \ldots, T\}$, which selects block $h$ in stage 1,  in the transmission state, reception state, and idle state respectively in the given frame.  When each node out of $n$ active nodes independently selects a block out of blocks $\{1, \ldots, t_T\}$ using the distribution in \eqref{EQ:pi}, let $u'(n, h)$ denote the probability that none of the nodes select a given block $h$ and $v'{(n, h)}$ denote the probability that exactly one node selects a given block $h$. So:
\begin{align}
u'{(n, h)} & =  \big(1 - {p'_h}\big)^{n}, \label{EQ:u'} \\
v'{(n, h)} & = n p'_h\big(1 - p'_h\big)^{{n} -1}.  \label{EQ:v'}
\end{align}

Recall that BP$_1$ and BP$_2$ denote the BPs broadcast by the BS after the end of stage 1 and stage 2 respectively, as shown in Fig.~\ref{Est_Window}. For $h \in \{1, \ldots, t_T\}$, let $d_h \in \{0,1\}$  be the value of the $h^{th}$ bit in BP$_1$. Recall from Section~\ref{3SS} that nodes of $\mathscr{T}_{1}$ participate in stage 1 and may participate in stage 2 (based on the corresponding bit value in BP$_1$), and nodes of $\mathscr{T}_{b}$, $b \ge 2$, participate in stage 1 and may participate in stage 3 (based on the corresponding bit values in BP$_1$ and BP$_2$). 

\paragraph{Expected Energy Consumption of a $\mathscr{T}_{1}$ Node}
Node $w_1$, upon choosing block $h$ using the distribution in \eqref{EQ:pi} in stage 1,  transmits symbol $\alpha$ in all ($T-1$) slots of that block (see Section~\ref{3SS}); hence, it consumes $(T-1)\gamma_\tau$ energy for transmission in this stage. Now, if $d_h = 1$ (respectively, $d_h = 0$), then node $w_1$ consumes $\gamma_\tau$ (respectively, 0) energy in stage 2 for transmission. The events in which $d_h = 1$ are: (a) At least one node from $\mathscr{N}_{1} \setminus \{w_1\}$ transmits in block $h$, and (b) At least one node each from $\mathscr{N}_{2}, \ldots, \mathscr{N}_{T}$ transmits and no node from $\mathscr{N}_{1} \setminus \{w_1\}$ transmits in block $h$. So, $P(\{d_h = 1\}) = Q'_{1} (h) + Q'_{2} (h)$, where $Q'_{1} (h)$ (respectively, $Q'_{2} (h)$) is the probability that event (a) (respectively, event (b)) occurs. Clearly, $Q'_{1} (h) = 1 - u'{(n_1 - 1, h)}$ (respectively, $Q'_{2} (h) = u'{(n_1 - 1, h)} \prod_{i=2}^{T}\left(1 - u'{(n_i, h)}\right)$). Hence, for a given $h$, the energy consumption of node $w_1$ in the transmission state is: $\mathscr{E}_{1, \tau}^{(h)}  = \left((T-1) + I_{\{d_h = 1\}}\right)\gamma_\tau$ and: 
\begin{align}
\label{eq_3SS_T}
E\left(\mathscr{E}_{1, \tau}^{(h)}\right) = \left( (T-1) + Q'_{1} (h) + Q'_{2} (h) \right)\gamma_\tau. 
\end{align}
Node $w_1$ reads all the slots of BP$_1$ (see Section~\ref{3SS}) and it consumes $\gamma_\rho$ energy in each slot. So:
\begin{equation}
\label{eq_3SS_R}
E\left(\mathscr{E}_{1, \rho}^{(h)}\right) = \left(\ceil{t_T/S_W}\right)\gamma_\rho.
\end{equation}
In the rest of the slots of phase 1, node $w_1$ is in the idle state. So:
\begin{align}
\label{eq_3SS_I}
E\left(\mathscr{E}_{1, \iota}^{(h)}\right) = \left( \Lambda_{I} - ((T-1) + Q'_{1} (h) + Q'_{2} (h)) - \ceil{t_T/S_W} \right)\gamma_\iota. 
\end{align}
 The total expected energy consumption of node $w_1$ is $E\left(\mathscr{E}_{1, \tau}^{(h)}\right) + E\left(\mathscr{E}_{1, \rho}^{(h)}\right) + E\left(\mathscr{E}_{1, \iota}^{(h)}\right)$, where $\mathscr{E}_{1, \tau}^{(h)}$, $\mathscr{E}_{1, \rho}^{(h)}$ and $\mathscr{E}_{1, \iota}^{(h)}$ are given by \eqref{eq_3SS_T}, \eqref{eq_3SS_R}, and \eqref{eq_3SS_I} respectively.

\paragraph{Expected Energy Consumption of a $\mathscr{T}_{b}$, $b \in \{2, \ldots, T\}$, Node}
Node $w_b$, upon choosing block $h$ using the distribution in \eqref{EQ:pi} in stage 1,  transmits symbol $\beta$ in only one slot of that block (see Section~\ref{3SS}); hence, it consumes $\gamma_\tau$ energy for transmission in this stage. Now, based on the  bit values corresponding to block $h$ in BP$_1$ and BP$_2$, node $w_b$ consumes $\gamma_\tau$ (respectively, 0) energy for transmission in stage 3  if both corresponding bits are 1 (respectively, at least one of them is 0). Hence:
\begin{align}
\label{eq_3SS_Tj}
E\left(\mathscr{E}_{b, \tau}^{(h)}\right) & = \left( 1 + Q''_{1} (h) \right)\gamma_\tau, 
\end{align}
where $Q''_{1} (h)$ is the probability that the event (c) occurs and (c) is the event that at least two nodes from $\mathscr{N}_{1}$ transmit in block $h$. Clearly,  $Q''_{1}(h) = 1 - u'{(n_1 ,h)} - v'{(n_1, h)}$.
Node $w_b$ first reads BP$_1$ (see Section~\ref{3SS}). If $d_h = 0$, then it does not read BP$_2$. Else, it reads only its corresponding slot of BP$_2$.  $d_h = 1$ iff event (c), (d), or (e) occurs, where (d) is the event that exactly one node from $\mathscr{N}_{1}$ transmits in block $h$ and at least one node from each of $\mathscr{N}_{2}, \ldots, \mathscr{N}_{b-1}, \mathscr{N}_{b+1}, \ldots, \mathscr{N}_{T}$ transmits in block $h$, and (e) is the event that at least two nodes each from $\mathscr{N}_{2}, \ldots, \mathscr{N}_{b-1}, \mathscr{N}_{b+1}, \ldots, \mathscr{N}_{T}$ transmit in block $h$, at least one node from $\mathscr{N}_{b} \setminus \{w_b\}$ transmits in block $h$ and no node from $\mathscr{N}_{1}$ transmits in block $h$. So, $P(\{d_h = 1\}) = Q''_{1} (h) + Q'''_{2} (h, b) + Q'''_{3} (h, b)$, where  $Q'''_{2} (h, b)$ (respectively, $Q'''_{3} (h, b)$) is the probability that event (d) (respectively, event (e)) occurs. Clearly:  
\begin{align}
Q'''_{2} (h, b) & = v'(n_1, h) \prod_{\substack{i=2 \\ i \ne b}}^{T}\left(1 - u'(n_i, h)\right), \label{eq_Q'''(2)} \\
Q'''_{3} (h, b) & = u'(n_1, h) \left(1 - u'(n_{b}-1, h)\right) \nonumber \\ 
		    & \prod_{\substack{i=2 \\ i \ne b}}^{T} \left(1 - u'(n_i, h) - v'(n_i, h)\right). \label{eq_Q'''(3)}
\end{align}
Now, $\mathscr{E}_{b, \rho}^{(h)}  = \left(\ceil{t_T/S_W} + I_{\{d_h = 1\}}\right)\gamma_\rho$ and: 
\begin{equation}
\label{eq_3SS_Rj}
E\left(\mathscr{E}_{b, \rho}^{(h)}\right) = \left(\ceil{t_T/S_W} + Q''_1 (h) + Q'''_{2} (h, b) + Q'''_{3} (h, b)\right) \gamma_\rho.
\end{equation}

In the rest of the slots of phase 1, node $w_b$ is  in the idle state (see Section~\ref{3SS}). So:
\begin{multline}
\label{eq_3SS_Ij}
E\left(\mathscr{E}_{b, \iota}^{(h)}\right) = \Big( \Lambda_{I} - (1 + Q''_{1} (h)) \\ - \left(\ceil{t_T/S_W} + Q''_1 (h) + Q'''_{2} (h, b) + Q'''_{3} (h, b)\right) \Big)\gamma_\iota
\end{multline}
The total expected energy consumption of node $w_b$ is: $E\left(\mathscr{E}_{b, \tau}^{(h)}\right) + E\left(\mathscr{E}_{b, \rho}^{(h)}\right) + E\left(\mathscr{E}_{b, \iota}^{(h)}\right)$, where $\mathscr{E}_{b, \tau}^{(h)}$, $\mathscr{E}_{b, \rho}^{(h)}$ and $\mathscr{E}_{b, \iota}^{(h)}$ are given by \eqref{eq_3SS_Tj},  \eqref{eq_3SS_Rj} and \eqref{eq_3SS_Ij} respectively.

\subsubsection{Expected Energy Consumption of a Node in 3-$SS$-$BB$}\label{Energy_3SSBB}
Expressions for the energy consumption of nodes of each type $b \in \{1, \ldots, T\}$ can be found by using a procedure similar to that in Section~\ref{SSSC:3SS:energy:consumption} with $\Lambda_{I}$ replaced with $\Lambda_{II}$, $t_T$ with $\ell$, $h \in \{1, \ldots, t_T\}$ with $i \in \{1, \ldots, \ell\}$, $u'(n, h)$ (see \eqref{EQ:u'}) with $u_b(n_b)$ (see \eqref{equnb}),  $v'(n, h)$ (see \eqref{EQ:v'}) with $v_b(n_b)$ (see \eqref{eqvnb}), and $p'_h$ (see \eqref{EQ:pi}) with $p_b$ (see \eqref{EQ:pb}) throughout. We omit the details for brevity. 

\subsubsection{Expected Energy Consumption of a Node in $T$-$Rep$-$BB$}\label{Energy_TRepBB}
Recall from Section~\ref{SubSec_Est} that each node of $\mathscr{T}_b$, $b \in \{1, \ldots, T\}$, transmits in one slot  (respectively, does not transmit in any slot) with probability $p_b$ (respectively, $1 - p_b$), where $p_b=\min \left({1,\frac{1.6\ell}{\tilde{n}_b}} \right)$ (see \eqref{EQ:pb}). So the expected energies consumed by an active node of $\mathscr{T}_{b}$ in the transmit, receive and idle states are 
$E\left(\mathscr{E}_{b, \tau}\right) = p_b \gamma_\tau$, $E\left(\mathscr{E}_{b, \rho}\right) =0$, and $E\left(\mathscr{E}_{b, \iota}\right) = (\ell -p_b) \gamma_\iota$ respectively.
Hence, for each $b \in \{1, \ldots, T\}$, the total energy consumed by an active node of $\mathscr{T}_{b}$ is:  
\begin{equation}
\label{eq_Energy_TRepBB}
E\left(\mathscr{E}_{b}\right) = p_b \gamma_\tau + (\ell -p_b) \gamma_\iota. 
\end{equation}

\subsubsection{Expected Energy Consumption of a Node in HSRC-1}
\label{SSSC:Energy_HSRC1}
Since in phase 1 of HSRC-1, 3-$SS$ is executed $M^{\prime}$ times and in phase 2,  either 3-$SS$-$BB$ or $T$-$Rep$-$BB$  is executed only once (see Section~\ref{EstScheme}), the total energy consumed by an active node of $\mathscr{T}_b$ under HSRC-1 is:
\begin{align}
E\left(\mathscr{E}_{b}^{\text{HSRC-1}}\right)  =  M' \left( E \left( E \left(  \mathscr{E}_{b, \tau}^{(h)} + \mathscr{E}_{b, \rho}^{(h)} + \mathscr{E}_{b, \iota}^{(h)}  \right) \right) \right)  + E \left(   \mathscr{E}_{b}\right),
\end{align}
where $E\left(\mathscr{E}_{b, \tau}^{(h)}\right)$, $E\left(\mathscr{E}_{b, \rho}^{(h)}\right)$ and $E\left(\mathscr{E}_{b, \iota}^{(h)}\right)$ are given by  \eqref{eq_3SS_T}, \eqref{eq_3SS_R}, and \eqref{eq_3SS_I} respectively if $b = 1$ and by \eqref{eq_3SS_Tj}, \eqref{eq_3SS_Rj}, and \eqref{eq_3SS_Ij} respectively if $b \in \{2, \ldots, T\}$. The outer expectation in the first term on the RHS is over the block number $h$, which is chosen using the distribution in \eqref{EQ:pi}. Also, $E\left(\mathscr{E}_{b}\right)$ is the energy consumed by an active node of $\mathscr{T}_{b}$ in phase 2 and its value is computed as explained in  Section~\ref{Energy_3SSBB} if 3-$SS$-$BB$ is used and using \eqref{eq_Energy_TRepBB} if $T$-$Rep$-$BB$ is used in phase 2.

\section{Simulations}\label{Simu}
We present simulation results in this section. Throughout, we assume that the parameter $S_W = 6$ and that the desired error probability  is $\delta=0.2$; hence, $M^{\prime} = 10$ (see Sections~\ref{SubSec_Est} and~\ref{EstScheme}). 

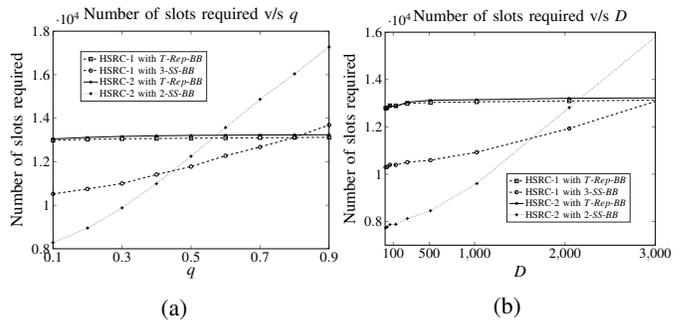
\begin{figure}
\centering
\begin{subfigure}{.5\textwidth}
\centering
\begin{adjustbox}{width = 1\columnwidth}
%
%
\begin{tikzpicture}

\begin{axis}[%
width=4.521in,
height=3.566in,
at={(0.758in,0.481in)},
scale only axis,
xmin=0.1,
xmax=0.9,
xtick={0.1, 0.3, 0.5, 0.7, 0.9},
yticklabel style = {font=\LARGE},
xticklabel style = {font=\LARGE},
xlabel style={font=\color{white!15!black}, font=\huge},
xlabel={$q$},
ymin=8000,
ymax=18000,
ytick={8000, 10000, 12000, 14000, 16000, 18000},
ylabel style={font=\color{white!15!black}, font=\huge},
ylabel={Number of slots required},
axis background/.style={fill=white},
title style={font=\color{white!15!black}, font=\huge},
title={Number of slots required v/s $q$},
legend style={font=\large, at={(0.1,0.678)}, anchor=south west, legend cell align=left, align=left, draw=white!15!black}
]
%
%

\addplot [color=black, dashed, mark=square, mark options={solid, black}]
  table[row sep=crcr]{%
0.1	12985.829\\
0.2	13030.857\\
0.3	13054.879\\
0.4	13070.183\\
0.5	13082.62\\
0.6	13093.508\\
0.7	13104.528\\
0.8	13114.007\\
0.9	13124.293\\
};
\addlegendentry{HSRC-1 with $T$-$Rep$-$BB$}

\addplot [color=black, dashed, mark=o, mark options={solid, black}]
  table[row sep=crcr]{%
0.1	10520.829\\
0.2	10748.857\\
0.3	10998.879\\
0.4	11411.183\\
0.5	11777.62\\
0.6	12274.508\\
0.7	12676.528\\
0.8	13123.007\\
0.9	13688.293\\
};
\addlegendentry{HSRC-1 with 3-$SS$-$BB$}


\addplot [color=black, mark=x, mark options={solid, black}]
  table[row sep=crcr]{%
0.1	13049.7146666667\\
0.2	13118.6146666667\\
0.3	13164.0316666667\\
0.4	13191.8526666667\\
0.5	13209.4356666667\\
0.6	13220.4686666667\\
0.7	13228.4146666667\\
0.8	13233.7926666667\\
0.9	13238.4656666667\\
};
\addlegendentry{HSRC-2 with $T$-$Rep$-$BB$}


\addplot [color=black, dotted, mark=+, mark options={solid, black}]
  table[row sep=crcr]{%
0.1	8281.71466666668\\
0.2	8951.61466666668\\
0.3	9877.03166666668\\
0.4	10989.8526666667\\
0.5	12251.4356666667\\
0.6	13571.4686666667\\
0.7	14868.4146666667\\
0.8	16037.7926666667\\
0.9	17274.4656666667\\
};
\addlegendentry{HSRC-2 with 2-$SS$-$BB$}

\end{axis}
\end{tikzpicture}%
\end{adjustbox}
\caption{}
\label{Sim_9q}
\end{subfigure}%
\begin{subfigure}{.5\textwidth}
\centering
\begin{adjustbox}{width = 1\columnwidth}
%
%
\begin{tikzpicture}

\begin{axis}[%
width=4.521in,
height=3.566in,
at={(0.758in,0.481in)},
scale only axis,
xmin=10,
xmax=3000,
xtick={100,500,1000,2000,3000},
yticklabel style = {font=\LARGE},
xticklabel style = {font=\LARGE},
xlabel style={font=\color{white!15!black}, font=\huge},
xlabel={$D$},
ymin=7000,
ymax=16000,
ytick={8000, 10000, 12000, 14000, 16000},
ylabel style={font=\color{white!15!black}, font=\huge},
ylabel={Number of slots required},
axis background/.style={fill=white},
title style={font=\color{white!15!black}, font=\huge},
title={Number of slots required v/s $D$},
legend style={font=\large, at={(0.45,0.1078)}, anchor=south west, legend cell align=left, align=left, draw=white!15!black}
]
%
%

\addplot [color=black, dashed, mark=square, mark options={solid, black}]
  table[row sep=crcr]{%
8	12762.082\\
16	12794.546\\
32	12806.319\\
64	12896.405\\
128	12885.171\\
256	12991.004\\
512	13030.136\\
1024	13054.974\\
2048	13094.866\\
4096	13167.28\\
};
\addlegendentry{HSRC-1 with $T$-$Rep$-$BB$}

\addplot [color=black, dashed, mark=o, mark options={solid, black}]
  table[row sep=crcr]{%
8	10255.082\\
16	10292.546\\
32	10304.319\\
64	10394.405\\
128	10391.171\\
256	10505.004\\
512	10591.136\\
1024	10926.974\\
2048	11931.866\\
4096	14391.28\\
};
\addlegendentry{HSRC-1 with 3-$SS$-$BB$}


\addplot [color=black, mark=x, mark options={solid, black}]
  table[row sep=crcr]{%
8	12680.2776666667\\
16	12746.1606666667\\
32	12784.7516666667\\
64	12877.0596666667\\
128	12865.9246666667\\
256	13040.9876666667\\
512	13122.7436666667\\
1024	13146.4766666667\\
2048	13206.7786666667\\
4096	13243.0326666667\\
};
\addlegendentry{HSRC-2 with $T$-$Rep$-$BB$}


\addplot [color=black, dotted, mark=+, mark options={solid, black}]
  table[row sep=crcr]{%
8	7665.27766666666\\
16	7731.16066666666\\
32	7769.75166666666\\
64	7869.05966666667\\
128	7881.92466666667\\
256	8125.98766666668\\
512	8454.74366666668\\
1024	9603.47666666668\\
2048	12813.7786666667\\
4096	19248.0326666667\\
};
\addlegendentry{HSRC-2 with 2-$SS$-$BB$}

\end{axis}
\end{tikzpicture}%
\end{adjustbox}
\caption{}
\label{Sim_9D}
\end{subfigure}
\caption{These plots show the average number of slots required by HSRC-1 with $T$-$Rep$-$BB$, HSRC-1 with 3-$SS$-$BB$, HSRC-2 with $T$-$Rep$-$BB$, and HSRC-2 with 2-$SS$-$BB$. The following parameters are used: $T$ = 4, $\epsilon$ = 0.03, $D$ = 1000  (in the left plot) and $q$ = 0.8 (in the right plot).}
\label{Sim_9Cases}
\end{figure}

\begin{figure}
\centering
\begin{subfigure}{.5\textwidth}
\centering
\begin{adjustbox}{width = 1\columnwidth}
%
%
\begin{tikzpicture}

\begin{axis}[%
width=4.521in,
height=3.566in,
at={(0.758in,0.481in)},
scale only axis,
xmin=500,
xmax=3000,
xtick={500, 1000, 1500, 2000, 2500, 3000},
yticklabel style = {font=\LARGE},
xticklabel style = {font=\LARGE},
xlabel style={font=\color{white!15!black}, font=\huge},
xlabel={$n_2$},
ymin=7000,
ymax=17000,
ytick={ 8000, 10000, 12000, 14000, 16000},
ylabel style={font=\color{white!15!black}, font=\huge},
ylabel={Number of slots required},
axis background/.style={fill=white},
title style={font=\color{white!15!black}, font=\huge},
title={Number of slots required v/s $n_2$},
legend style={font=\Large, at={(0.056,0.602)}, anchor=south west, legend cell align=left, align=left, draw=white!15!black}
]
\addplot [color=black]
  table[row sep=crcr]{%
500	12036\\
600	12036\\
700	12036\\
800	12036\\
900	12036\\
1000	12036\\
1100	12036\\
1200	12036\\
1300	12036\\
1400	12036\\
1500	12036\\
1600	12036\\
1700	12036\\
1800	12036\\
1900	12036\\
2000	12036\\
2100	12036\\
2200	12036\\
2300	12036\\
2400	12036\\
2500	12036\\
2600	12036\\
2700	12036\\
2800	12036\\
2900	12036\\
3000	12036\\
};
\addlegendentry{$T$-$Rep$-$BB$}

\addplot [color=black, dotted, mark=*, mark options={solid, black}]
  table[row sep=crcr]{%
500	9781.3\\
600	9782.4\\
700	9784.5\\
800	9801.7\\
900	9789\\
1000	9790.7\\
1100	9788\\
1200	9786.6\\
1300	9780.2\\
1400	9809.5\\
1500	9774.1\\
1600	9799.9\\
1700	9800.5\\
1800	9783.6\\
1900	9788.6\\
2000	9798.1\\
2100	9787.8\\
2200	9790.8\\
2300	9783.9\\
2400	9792.5\\
2500	9797.7\\
2600	9789.2\\
2700	9799.5\\
2800	9801\\
2900	9790.7\\
3000	9804.4\\
};
\addlegendentry{3-$SS$-$BB$; $n_1 = n_3 = n_4 = 500$}

\addplot [color=black, mark=diamond, mark options={solid, black}]
  table[row sep=crcr]{%
500	7836\\
600	7923.7\\
700	8021.3\\
800	8132.4\\
900	8244.5\\
1000	8351.5\\
1100	8440.3\\
1200	8571\\
1300	8697.2\\
1400	8816.4\\
1500	8929.7\\
1600	9063.4\\
1700	9199.3\\
1800	9313.3\\
1900	9432.7\\
2000	9569.7\\
2100	9731.1\\
2200	9850.8\\
2300	9989.2\\
2400	10104.2\\
2500	10246.6\\
2600	10395.2\\
2700	10529.2\\
2800	10645.3\\
2900	10773.7\\
3000	10939.4\\
};
\addlegendentry{2-$SS$-$BB$; $n_1 = n_3 = n_4 = 500$}

\addplot [color=black, dotted, mark=o, mark options={solid, black}]
  table[row sep=crcr]{%
500	10203.4\\
600	10180\\
700	10217.7\\
800	10193.7\\
900	10198\\
1000	10199.3\\
1100	10193.7\\
1200	10221.8\\
1300	10228.4\\
1400	10192.8\\
1500	10218.7\\
1600	10241.5\\
1700	10230.5\\
1800	10224\\
1900	10214.3\\
2000	10221.7\\
2100	10224.5\\
2200	10217.8\\
2300	10218.7\\
2400	10223.5\\
2500	10213.4\\
2600	10243.8\\
2700	10223.8\\
2800	10232.9\\
2900	10234.4\\
3000	10220.6\\
};
\addlegendentry{3-$SS$-$BB$; $n_1 = n_3 = n_4 = 1000$}

\addplot [color=black, mark=asterisk, mark options={solid, black}]
  table[row sep=crcr]{%
500	8728.5\\
600	8845.7\\
700	8958.1\\
800	9091.9\\
900	9223.7\\
1000	9348.7\\
1100	9501.3\\
1200	9647.7\\
1300	9780\\
1400	9867.9\\
1500	10030.7\\
1600	10204.7\\
1700	10327.5\\
1800	10456.8\\
1900	10603\\
2000	10748.1\\
2100	10869\\
2200	10981.7\\
2300	11160.5\\
2400	11258.2\\
2500	11443.8\\
2600	11570.2\\
2700	11677.8\\
2800	11824.9\\
2900	11972.8\\
3000	12106.7\\
};
\addlegendentry{2-$SS$-$BB$; $n_1 = n_3 = n_4 = 1000$}

\end{axis}
\end{tikzpicture}%
\end{adjustbox}
\caption{}
\label{Sim_n2_T4}
\end{subfigure}%
\begin{subfigure}{.5\textwidth}
\centering
\begin{adjustbox}{width = 1\columnwidth}
%
%
\begin{tikzpicture}

\begin{axis}[%
width=4.521in,
height=3.566in,
at={(0.758in,0.481in)},
scale only axis,
xmin=500,
xmax=3000,
xtick={500, 1000, 1500, 2000, 2500, 3000},
yticklabel style = {font=\LARGE},
xticklabel style = {font=\LARGE},
xlabel style={font=\color{white!15!black}, font=\huge},
xlabel={$n_2$},
ymin=8000,
ymax=20000,
ytick={ 8000, 10000, 12000, 14000, 16000, 18000, 20000},
ylabel style={font=\color{white!15!black}, font=\huge},
ylabel={Number of slots required},
axis background/.style={fill=white},
title style={font=\color{white!15!black}, font=\huge},
title={Number of slots required v/s $n_2$},
legend style={font=\Large, at={(0.026,0.672)}, anchor=south west, legend cell align=left, align=left, draw=white!15!black}
]
\addplot [color=black]
  table[row sep=crcr]{%
500	15045\\
600	15045\\
700	15045\\
800	15045\\
900	15045\\
1000	15045\\
1100	15045\\
1200	15045\\
1300	15045\\
1400	15045\\
1500	15045\\
1600	15045\\
1700	15045\\
1800	15045\\
1900	15045\\
2000	15045\\
2100	15045\\
2200	15045\\
2300	15045\\
2400	15045\\
2500	15045\\
2600	15045\\
2700	15045\\
2800	15045\\
2900	15045\\
3000	15045\\
};
\addlegendentry{$T$-$Rep$-$BB$}

\addplot [color=black, dotted, mark=*, mark options={solid, black}]
  table[row sep=crcr]{%
500	12824.4\\
600	12834.5\\
700	12830.5\\
800	12831\\
900	12834.8\\
1000	12825.6\\
1100	12818.5\\
1200	12854.3\\
1300	12829.6\\
1400	12833.5\\
1500	12827\\
1600	12842.6\\
1700	12827.9\\
1800	12852.1\\
1900	12849.7\\
2000	12827.1\\
2100	12836.7\\
2200	12828.2\\
2300	12839.3\\
2400	12836.1\\
2500	12834.1\\
2600	12827.4\\
2700	12818.8\\
2800	12838\\
2900	12821.7\\
3000	12838\\
};
\addlegendentry{3-$SS$-$BB$; $n_1 = n_3 = n_4 = n_5 = 500$}

\addplot [color=black, mark=diamond, mark options={solid, black}]
  table[row sep=crcr]{%
500	9183\\
600	9340.4\\
700	9435.5\\
800	9598.3\\
900	9733.7\\
1000	9920.1\\
1100	10059.9\\
1200	10219\\
1300	10368.2\\
1400	10542\\
1500	10717.8\\
1600	10862.8\\
1700	11034.3\\
1800	11214.1\\
1900	11395.6\\
2000	11539.5\\
2100	11689.8\\
2200	11871.6\\
2300	12034.1\\
2400	12232.3\\
2500	12420.2\\
2600	12545.2\\
2700	12728.7\\
2800	12938.5\\
2900	13036.7\\
3000	13225.5\\
};
\addlegendentry{2-$SS$-$BB$; $n_1 = n_3 = n_4 = n_5 = 500$}

\addplot [color=black, dotted, mark=o, mark options={solid, black}]
  table[row sep=crcr]{%
500	13315.3\\
600	13328.6\\
700	13349.4\\
800	13339.6\\
900	13348.1\\
1000	13380.3\\
1100	13331.9\\
1200	13342.6\\
1300	13333.6\\
1400	13344.1\\
1500	13349.7\\
1600	13332.9\\
1700	13353.8\\
1800	13342.1\\
1900	13332\\
2000	13359\\
2100	13331.3\\
2200	13325.2\\
2300	13343.7\\
2400	13346.8\\
2500	13333.8\\
2600	13355.6\\
2700	13331.5\\
2800	13329.6\\
2900	13338.2\\
3000	13363.7\\
};
\addlegendentry{3-$SS$-$BB$; $n_1 = n_3 = n_4 = n_5 = 1000$}

\addplot [color=black, mark=asterisk, mark options={solid, black}]
  table[row sep=crcr]{%
500	11093.8\\
600	11267.5\\
700	11444.6\\
800	11572.4\\
900	11776.8\\
1000	11924.4\\
1100	12086.6\\
1200	12224.5\\
1300	12430.9\\
1400	12605.1\\
1500	12772\\
1600	12929.4\\
1700	13136.4\\
1800	13259.7\\
1900	13461.2\\
2000	13594.3\\
2100	13767\\
2200	13896.4\\
2300	14057.8\\
2400	14228.9\\
2500	14356.2\\
2600	14526.9\\
2700	14719.2\\
2800	14846.9\\
2900	14976.1\\
3000	15178.5\\
};
\addlegendentry{2-$SS$-$BB$; $n_1 = n_3 = n_4 = n_5 = 1000$}

\end{axis}
\end{tikzpicture}%
\end{adjustbox}
\caption{}
\label{Sim_n2_T5}
\end{subfigure}
\caption{These plots show the average number of slots required in phase 2 of the proposed estimation protocols versus $n_2$ when the 3-$SS$-$BB$, 2-$SS$-$BB$, and $T$-$Rep$-$BB$ methods are used in phase 2. The following parameters are used: $\epsilon$ = 0.03, $\ell = 3009$, $T$ = 4 (in Fig.~\ref{Sim_n2_T4}) and $T$ = 5 (in Fig.~\ref{Sim_n2_T5}).}
\label{Sim5}
\end{figure}
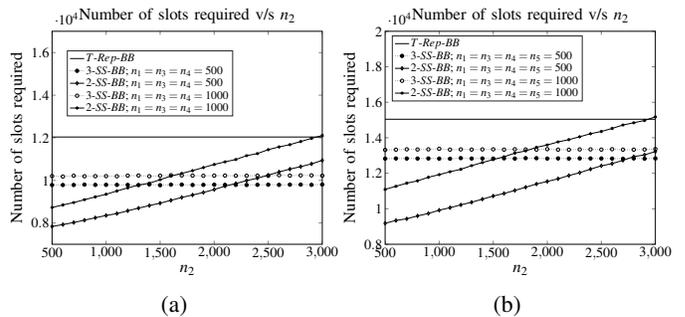

\begin{figure}
\centering
\begin{subfigure}{.5\textwidth}
\centering
\begin{adjustbox}{width = 1\columnwidth}
%
%
\begin{tikzpicture}

\begin{axis}[%
width=4.521in,
height=3.566in,
at={(0.758in,0.481in)},
scale only axis,
xmin=500,
xmax=3000,
xtick={500, 1000, 1500, 2000, 2500, 3000},
yticklabel style = {font=\LARGE},
xticklabel style = {font=\LARGE},
xlabel style={font=\color{white!15!black}, font=\huge},
xlabel={$n_1$},
ymin=7000,
ymax=17000,
ytick={ 8000, 10000, 12000, 14000, 16000},
ylabel style={font=\color{white!15!black}, font=\huge},
ylabel={Number of slots required},
axis background/.style={fill=white},
title style={font=\color{white!15!black}, font=\huge},
title={Number of slots required v/s $n_1$},
legend style={font=\Large, at={(0.056,0.602)}, anchor=south west, legend cell align=left, align=left, draw=white!15!black}
]
\addplot [color=black]
  table[row sep=crcr]{%
500	12036\\
600	12036\\
700	12036\\
800	12036\\
900	12036\\
1000	12036\\
1100	12036\\
1200	12036\\
1300	12036\\
1400	12036\\
1500	12036\\
1600	12036\\
1700	12036\\
1800	12036\\
1900	12036\\
2000	12036\\
2100	12036\\
2200	12036\\
2300	12036\\
2400	12036\\
2500	12036\\
2600	12036\\
2700	12036\\
2800	12036\\
2900	12036\\
3000	12036\\
};
\addlegendentry{$T$-$Rep$-$BB$}

\addplot [color=black, dotted, mark=*, mark options={solid, black}]
  table[row sep=crcr]{%
500	9775.5\\
600	9858.1\\
700	9918.5\\
800	10005.7\\
900	10086.5\\
1000	10201.9\\
1100	10298.9\\
1200	10407.9\\
1300	10524\\
1400	10654.3\\
1500	10747\\
1600	10891.6\\
1700	11029.3\\
1800	11164.4\\
1900	11330.2\\
2000	11466.6\\
2100	11617.8\\
2200	11757.3\\
2300	11903.8\\
2400	12045.5\\
2500	12158.4\\
2600	12319.2\\
2700	12513.7\\
2800	12659.9\\
2900	12817.9\\
3000	12959.2\\
};
\addlegendentry{3-$SS$-$BB$; $n_2 = n_3 = n_4 = 500$}

\addplot [color=black, mark=diamond, mark options={solid, black}]
  table[row sep=crcr]{%
500	7820.6\\
600	7865.8\\
700	7877.9\\
800	7913.6\\
900	7921.7\\
1000	7942.7\\
1100	7988.8\\
1200	8000.1\\
1300	8054.2\\
1400	8054.2\\
1500	8079.2\\
1600	8097.6\\
1700	8120\\
1800	8155.7\\
1900	8174\\
2000	8193\\
2100	8198\\
2200	8234.2\\
2300	8265\\
2400	8261.6\\
2500	8287.7\\
2600	8332.4\\
2700	8332.4\\
2800	8347.2\\
2900	8384\\
3000	8402.2\\
};
\addlegendentry{2-$SS$-$BB$; $n_2 = n_3 = n_4 = 500$}

\addplot [color=black, dotted, mark=o, mark options={solid, black}]
  table[row sep=crcr]{%
500	9798.5\\
600	9867.5\\
700	9945.5\\
800	10008.5\\
900	10121.7\\
1000	10215.9\\
1100	10322.9\\
1200	10435.2\\
1300	10545.7\\
1400	10671.1\\
1500	10773.6\\
1600	10907.1\\
1700	11047.7\\
1800	11186\\
1900	11324.3\\
2000	11492.3\\
2100	11608\\
2200	11785.5\\
2300	11919.6\\
2400	12034.2\\
2500	12174.1\\
2600	12364.6\\
2700	12523.5\\
2800	12672.6\\
2900	12866.5\\
3000	13007.2\\
};
\addlegendentry{3-$SS$-$BB$; $n_2 = n_3 = n_4 = 1000$}

\addplot [color=black, mark=asterisk, mark options={solid, black}]
  table[row sep=crcr]{%
500	9102.3\\
600	9165\\
700	9207.2\\
800	9259.8\\
900	9323.8\\
1000	9391.5\\
1100	9410.2\\
1200	9488.9\\
1300	9522.4\\
1400	9581.4\\
1500	9647\\
1600	9687.8\\
1700	9717\\
1800	9783.5\\
1900	9801.2\\
2000	9914.4\\
2100	9885.5\\
2200	9953.1\\
2300	9991.9\\
2400	10064.7\\
2500	10074\\
2600	10137\\
2700	10193.3\\
2800	10228.5\\
2900	10284.4\\
3000	10331.9\\
};
\addlegendentry{2-$SS$-$BB$; $n_2 = n_3 = n_4 = 1000$}

\end{axis}
\end{tikzpicture}%
\end{adjustbox}
\caption{}
\label{Sim_n1_T4}
\end{subfigure}%
\begin{subfigure}{.5\textwidth}
\centering
\begin{adjustbox}{width = 1\columnwidth}
%
%
\begin{tikzpicture}

\begin{axis}[%
width=4.521in,
height=3.566in,
at={(0.758in,0.481in)},
scale only axis,
xmin=500,
xmax=3000,
xtick={500, 1000, 1500, 2000, 2500, 3000},
yticklabel style = {font=\LARGE},
xticklabel style = {font=\LARGE},
xlabel style={font=\color{white!15!black}, font=\huge},
xlabel={$n_1$},
ymin=8000,
ymax=20000,
ytick={ 8000, 10000, 12000, 14000, 16000, 18000, 20000},
ylabel style={font=\color{white!15!black}, font=\huge},
ylabel={Number of slots required},
axis background/.style={fill=white},
title style={font=\color{white!15!black}, font=\huge},
title={Number of slots required v/s $n_1$},
legend style={font=\Large, at={(0.026,0.672)}, anchor=south west, legend cell align=left, align=left, draw=white!15!black}
]
\addplot [color=black]
  table[row sep=crcr]{%
500	15045\\
600	15045\\
700	15045\\
800	15045\\
900	15045\\
1000	15045\\
1100	15045\\
1200	15045\\
1300	15045\\
1400	15045\\
1500	15045\\
1600	15045\\
1700	15045\\
1800	15045\\
1900	15045\\
2000	15045\\
2100	15045\\
2200	15045\\
2300	15045\\
2400	15045\\
2500	15045\\
2600	15045\\
2700	15045\\
2800	15045\\
2900	15045\\
3000	15045\\
};
\addlegendentry{$T$-$Rep$-$BB$}

\addplot [color=black, dotted, mark=*, mark options={solid, black}]
  table[row sep=crcr]{%
500	12828.5\\
600	12899.8\\
700	13002.8\\
800	13124.9\\
900	13216.7\\
1000	13366.9\\
1100	13480.7\\
1200	13589.5\\
1300	13787.5\\
1400	13887.3\\
1500	14066.3\\
1600	14211.5\\
1700	14361.1\\
1800	14520.1\\
1900	14723.4\\
2000	14873.5\\
2100	15070.9\\
2200	15284.4\\
2300	15452.2\\
2400	15614.8\\
2500	15808.5\\
2600	15968\\
2700	16177.6\\
2800	16377.2\\
2900	16540.5\\
3000	16714.9\\
};
\addlegendentry{3-$SS$-$BB$; $n_2 = n_3 = n_4 = n_5 = 500$}

\addplot [color=black, mark=diamond, mark options={solid, black}]
  table[row sep=crcr]{%
500	9165.4\\
600	9245.2\\
700	9262.3\\
800	9306.3\\
900	9358.8\\
1000	9396\\
1100	9444.6\\
1200	9475.3\\
1300	9491.5\\
1400	9526.8\\
1500	9596.1\\
1600	9617.7\\
1700	9676.7\\
1800	9710.6\\
1900	9746.2\\
2000	9750.1\\
2100	9830.7\\
2200	9846.1\\
2300	9864.2\\
2400	9879\\
2500	9938.1\\
2600	10010.1\\
2700	10023.4\\
2800	10022.8\\
2900	10029.3\\
3000	10087.4\\
};
\addlegendentry{2-$SS$-$BB$; $n_2 = n_3 = n_4 = n_5 = 500$}

\addplot [color=black, dotted, mark=o, mark options={solid, black}]
  table[row sep=crcr]{%
500	12826.7\\
600	12917.8\\
700	13024.7\\
800	13106\\
900	13217.3\\
1000	13338.5\\
1100	13478.9\\
1200	13610.4\\
1300	13773.1\\
1400	13884.5\\
1500	14042.7\\
1600	14213.8\\
1700	14374.1\\
1800	14535.9\\
1900	14725\\
2000	14851.1\\
2100	15078.1\\
2200	15270.9\\
2300	15405\\
2400	15651\\
2500	15811.2\\
2600	15983.4\\
2700	16203.5\\
2800	16360.7\\
2900	16578.8\\
3000	16772.9\\
};
\addlegendentry{3-$SS$-$BB$; $n_2 = n_3 = n_4 = n_5 = 1000$}

\addplot [color=black, mark=asterisk, mark options={solid, black}]
  table[row sep=crcr]{%
500	11558\\
600	11635.7\\
700	11706\\
800	11800.8\\
900	11885.1\\
1000	11920.6\\
1100	12014.3\\
1200	12070.9\\
1300	12163.8\\
1400	12185.2\\
1500	12295.9\\
1600	12336.6\\
1700	12421.7\\
1800	12478.8\\
1900	12578.4\\
2000	12626.2\\
2100	12705\\
2200	12764.9\\
2300	12822.7\\
2400	12876.4\\
2500	12903.9\\
2600	13009.4\\
2700	13044.1\\
2800	13156.9\\
2900	13205.1\\
3000	13265.4\\
};
\addlegendentry{2-$SS$-$BB$; $n_2 = n_3 = n_4 = n_5 = 1000$}

\end{axis}
\end{tikzpicture}%
\end{adjustbox}
\caption{}
\label{Sim_n1_T5}
\end{subfigure}
\caption{These plots show the average number of slots required in phase 2 of the proposed estimation protocols versus $n_1$ when the 3-$SS$-$BB$, 2-$SS$-$BB$, and $T$-$Rep$-$BB$ methods are used in phase 2. The following parameters are used: $\epsilon$ = 0.03, $\ell = 3009$, $T$ = 4 (in Fig.~\ref{Sim_n1_T4}) and $T$ = 5 (in Fig.~\ref{Sim_n1_T5}).}
\label{Sim4}
\end{figure}
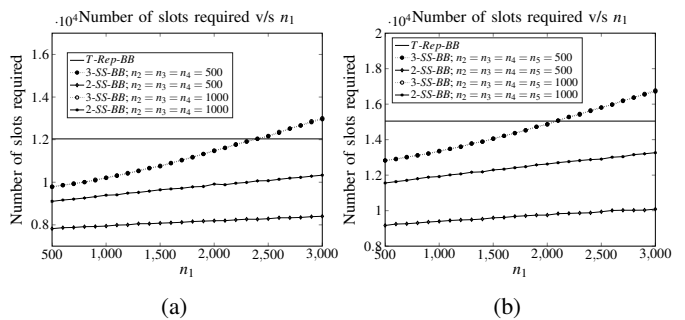

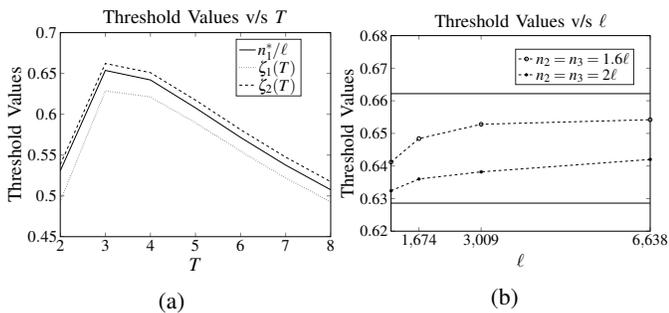
\begin{figure}
\centering
\begin{subfigure}{.5\textwidth}
\centering
\begin{adjustbox}{width = 1\columnwidth}
%
%

\begin{tikzpicture}

\begin{axis}[%
width=4.521in,
height=3.566in,
at={(0.758in,0.481in)},
xmin=2,
xmax=8,
xtick={2, 3, 4, 5, 6, 7, 8},
yticklabel style = {font=\LARGE},
xticklabel style = {font=\LARGE},
xlabel style={font=\color{white!15!black}, font=\huge},
xlabel={$T$},
ymin=0.45,
ymax=0.7,
ylabel style={font=\color{white!15!black}, font=\huge},
ylabel={Threshold Values},
axis background/.style={fill=white},
title style={font=\color{white!15!black}, font=\huge},
title={Threshold Values v/s $T$},
legend style={font=\LARGE, at={(0.65,0.689)}, anchor=south west, legend cell align=left, align=left, draw=white!15!black}
]
\addplot [color=black]
  table[row sep=crcr]{%
2	0.530741110003323\\
3	0.653705550016617\\
4	0.642073778664008\\
5	0.607843137254902\\
6	0.571618477899634\\
7	0.537720172814889\\
8	0.507477567298106\\
};
\addlegendentry{$n_1^*/\ell$}

\addplot [color=black, dotted]
  table[row sep=crcr]{%
2	0.4932\\
3	0.6286\\
4	0.6213\\
5	0.5897\\
6	0.5548\\
7	0.522\\
8	0.4926\\
};
\addlegendentry{$\zeta_1(T)$}

\addplot [color=black, dashed]
  table[row sep=crcr]{%
2	0.5384\\
3	0.6622\\
4	0.651\\
5	0.6173\\
6	0.5812\\
7	0.5475\\
8	0.5174\\
};
\addlegendentry{$\zeta_2(T)$}

\end{axis}
\end{tikzpicture}%
\end{adjustbox}
\caption{}
\label{x_star_fig}
\end{subfigure}%
\begin{subfigure}{.5\textwidth}
\centering
\begin{adjustbox}{width = 1\columnwidth}
%
%
\begin{tikzpicture}

\begin{axis}[%
width=4.521in,
height=3.566in,
at={(0.758in,0.481in)},
xmin=1075,
xmax=6638,
xtick={1674,3009,6638},
yticklabel style = {font=\LARGE},
xticklabel style = {font=\LARGE},
xlabel style={font=\color{white!15!black}, font=\huge},
xlabel={$\ell$},
ymin=0.62,
ymax=0.68,
ylabel style={font=\color{white!15!black}, font=\huge},
ylabel={Threshold Values},
axis background/.style={fill=white},
title style={font=\color{white!15!black}, font=\huge},
title={Threshold Values v/s $\ell$},
legend style={font=\LARGE, at={(0.48,0.749)}, anchor=south west, legend cell align=left, align=left, draw=white!15!black}
]
\addplot [color=black, dashed, mark=o, mark options={solid, black}]
  table[row sep=crcr]{%
1075	0.6412\\
1674	0.6484\\
3009	0.6528\\
6638	0.6542\\
};
\addlegendentry{$n_2 = n_3 = 1.6\ell$}

\addplot [color=black, dashed, mark=asterisk, mark options={solid, black}]
  table[row sep=crcr]{%
1075	0.6324\\
1674	0.636\\
3009	0.6382\\
6638	0.642\\
};
\addlegendentry{$n_2 = n_3 = 2\ell$}

\addplot [color=black, forget plot]
  table[row sep=crcr]{%
1075	0.6286\\
1674	0.6286\\
3009	0.6286\\
6638	0.6286\\
};
\addplot [color=black, forget plot]
  table[row sep=crcr]{%
1075	0.6622\\
1674	0.6622\\
3009	0.6622\\
6638	0.6622\\
};
\end{axis}
\end{tikzpicture}%
\end{adjustbox}
\caption{}
\label{Thr_vs_l}
\end{subfigure}
\caption{In the plot in (a), the parameter values $\ell = 3009$, $\delta = 0.2$, and $\epsilon = 0.03$ are used. The plot in (b) shows $n_1^*/\ell$ versus $\ell$ and the two bounds $0.6286$ and $0.6622$ for the case $T=3$. The following parameters are used: $\epsilon = [0.02, 0.03, 0.04, 0.05]$ and corresponding $\ell = [6638, 3009, 1674, 1075]$~\cite{zhou2016understanding}.}
\label{sim_3d1}
\end{figure}

Let $D$ be the total number of nodes of each type and suppose each node is active with probability $q$ in a frame. Recall that in phase 2 of HSRC-1 (respectively, HSRC-2), either $T$-$Rep$-$BB$ or 3-$SS$-$BB$ (respectively, 2-$SS$-$BB$)  is used. Figs.~\ref{Sim_9q} and~\ref{Sim_9D} show the average number of slots required to execute HSRC-1 with $T$-$Rep$-$BB$, HSRC-1 with 3-$SS$-$BB$, HSRC-2 with $T$-$Rep$-$BB$, and HSRC-2 with 2-$SS$-$BB$ versus $q$ and $D$ respectively. 
From \Figref{Sim_9q} (respectively, \Figref{Sim_9D}), we can observe that from $q = 0.1$ to $0.45$ (respectively, $D = 100$ to $1700$), HSRC-2 with 2-$SS$-$BB$ outperforms the other schemes, from $q = 0.45$ to $0.8$ (respectively, $D = 1700$ to $3000$), HSRC-1 with 3-$SS$-$BB$ outperforms the other schemes and for $q \ge 0.8 $ (respectively, $D \ge 3000$), HSRC-1 with $T$-$Rep$-$BB$ and  HSRC-2 with $T$-$Rep$-$BB$ outperform the other schemes. These results show that for sufficiently low values of $q$ (respectively, $D$),  both HSRC-1 with 3-$SS$-$BB$ and HSRC-2 with 2-$SS$-$BB$ outperform  HSRC-1 with $T$-$Rep$-$BB$ as well as  HSRC-2 with $T$-$Rep$-$BB$. Intuitively, this is because when $q$ or $D$ is low, only a few nodes are active, and hence only a small number of collisions occur in stage 1 and/ or stage 2 of phase 2 of HSRC-1 with 3-$SS$-$BB$ and HSRC-2 with 2-$SS$-$BB$.

Fig.~\ref{Sim_n2_T4} (respectively, Fig.~\ref{Sim_n2_T5}) shows the number of slots required in phase 2 of the proposed estimation protocols when 3-$SS$-$BB$, 2-$SS$-$BB$, and $T$-$Rep$-$BB$ are used versus $n_2$ for $T = 4$ (respectively, $T=5$) and two different pairs of values of $n_1$, $n_3$, and $n_4$ (respectively, $n_1$, $n_3$, $n_4$, and $n_5$). It can be seen that for each set of values of $n_1$, $n_3$, and $n_4$  (respectively, $n_1$, $n_3$, $n_4$, and $n_5$), the number of slots required by 3-$SS$-$BB$ remain approximately the same as $n_2$ changes; this is because in 3-$SS$-$BB$, a $\mathscr{T}_1$ node that selects a block $B_i$ in stage 1 transmits symbol $\alpha$ in \emph{all $(T-1)$} slots of the block, whereas a $\mathscr{T}_b$, $b \in \{2, \ldots, T\}$, node that selects a block $B_i$ in stage 1 transmits symbol $\beta$ in \emph{only one slot} and does not transmit in the other slots of block $B_i$, i.e., it can cause a collision in only one slot (see Fig.~\ref{Sym_Combo3}, Sections~\ref{GLOBECOM} and~\ref{EstScheme}). On the other hand, the number of slots required by 2-$SS$-$BB$ increases significantly in $n_2$. This is because in 2-$SS$-$BB$, $\mathscr{T}_2$ nodes use the symbol combination $\alpha, \alpha, 0, \ldots, 0$ for transmission, i.e., they can cause collisions in two slots (see Fig.~\ref{Sym_Combo}, Sections~\ref{GLOBECOM} and~\ref{EstScheme}); so the number of collisions in stage 1 significantly increases when $n_2$ increases.  Also, in both Fig.~\ref{Sim_n2_T4} and Fig.~\ref{Sim_n2_T5}, when $n_1$, $n_3$, and $n_4$  (respectively, $n_1$, $n_3$, $n_4$, and $n_5$) increase, the number of slots required by both  3-$SS$-$BB$ and 2-$SS$-$BB$ increase; again, this is because the number of collisions increases.

Fig.~\ref{Sim_n1_T4} (respectively, Fig.~\ref{Sim_n1_T5}) shows the number of slots required in phase 2 of the proposed estimation protocols when 3-$SS$-$BB$, 2-$SS$-$BB$, and $T$-$Rep$-$BB$  are used versus $n_1$ for $T = 4$ (respectively, $T=5$) and two different pairs of values of $n_2$ to $n_4$ (respectively, $n_2$ to $n_5$). It can be seen that for each set of values of $n_2$ to $n_4$  (respectively, $n_2$ to $n_5$), the number of slots required by both 3-$SS$-$BB$ and 2-$SS$-$BB$ increases in $n_1$; this is because the number of collisions in stage 1 increases.  Also, when $n_2$ to $n_4$  (respectively, $n_2$ to $n_5$) increase, the number of slots required by 2-$SS$-$BB$ increases; again, this is because the number of collisions increases. However, when $n_2$ to $n_4$  (respectively, $n_2$ to $n_5$) increase, the number of slots required by 3-$SS$-$BB$ remain almost unchanged; this is due to the reasons explained in the previous paragraph.

Let $n_1^*$ be the value of $\tilde{n}_1$ for which 3-$SS$-$BB$ and $T$-$Rep$-$BB$ require equal numbers of slots to execute on average in phase 2 of HSRC-1. Note that the value of $n_1^*$ can be obtained by using a plot such as Fig.~\ref{Sim_n1_T4} and noting the value of $n_1$ at which the curve for 3-$SS$-$BB$ intersects the horizontal line corresponding to $T$-$Rep$-$BB$. Fig.~\ref{x_star_fig} shows $n_1^*/\ell$, $\zeta_1(T)$, and $\zeta_2(T)$  for different values of $T$.\footnote{Methods to obtain $\zeta_1(T)$ and $\zeta_2(T)$ for different values of $T$ are provided in Remark~\ref{zeta_remark}.} From the figure, it is clear that $n_1^*/\ell$ lies between $\zeta_1(T)$ and $\zeta_2(T)$, $\forall T$, which is consistent with the analysis in Section~\ref{Boundary}.  Next, for $T=3$, 
Fig.~\ref{Thr_vs_l} shows a plot of $n_1^*/\ell$ versus $\ell$ for two different pairs of values of $n_2$ and $n_3$. Again, it can be seen that $\zeta_1 (3) = 0.6286 < n_1^*/\ell < \zeta_2 (3) = 0.6622$ for all values considered.\footnote{The values of $\zeta_1 (3) = 0.6286$ and $\zeta_2 (3) = 0.6622$ can be found either by observing Fig.~\ref{x_star_fig} or by using the methods provided in Remark~\ref{zeta_remark}.}  
From Section~\ref{Boundary}, we see that when $\tilde{n}_1 \leq 0.6286 \ell$ (respectively, $\tilde{n}_1 \geq 0.6622 \ell$), 3-$SS$-$BB$ takes less (respectively, more) time than $T$-$Rep$-$BB$. 
Also, by using a plot such as Fig.~\ref{Thr_vs_l}, we can find out $n_1^*$, using which we can in turn find out, for given values of $\tilde{n}_1$, $\tilde{n}_2$ and $\tilde{n}_3$, whether  using 3-$SS$-$BB$ or $T$-$Rep$-$BB$ would take fewer slots in phase 2 of HSRC-1 in practice-- note that if $\tilde{n}_1 < n_1^*$ (respectively, $\tilde{n}_1 > n_1^*$), then 3-$SS$-$BB$ (respectively,  $T$-$Rep$-$BB$) would take fewer slots.  

\begin{figure}
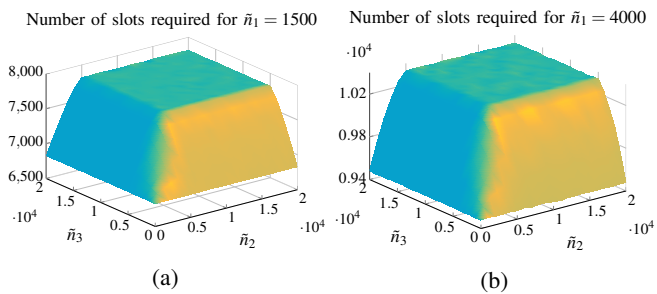

	\centering
	\begin{subfigure}{0.48\textwidth}
		\centering
		\begin{adjustbox}{width = \columnwidth}
		\input{Slots_Required_n1_1500_new}
		\end{adjustbox}
		\caption{}
		\label{Sim1}
	\end{subfigure}
	\begin{subfigure}{0.48\textwidth}
		\centering
		\begin{adjustbox}{width = \columnwidth}
		\input{Slots_Required_n1_4000_new}
		\end{adjustbox}
		\caption{}
		\label{Sim2}
	\end{subfigure}
	\caption{These plots show the average numbers of slots required in phase 2 of  HSRC-1 with 3-$SS$-$BB$ for $\tilde{n}_1 = 1500$ and $\tilde{n}_1 = 4000$. The following parameters are used: $T=3$, $\epsilon$ = 0.03, and $\ell = 3009$.}
	\label{sim_3d}
\end{figure}

In Figs.~\ref{Sim1} and~\ref{Sim2}, the average numbers of slots required in  phase 2  of HSRC-1 with 3-$SS$-$BB$ are plotted versus $\tilde{n}_2$ and $\tilde{n}_3$ for $\tilde{n}_1 = 1500$ and $\tilde{n}_1 = 4000$ respectively. 
It can be seen that in Fig.~\ref{Sim1}, for all the values of $\tilde{n}_2$ and $\tilde{n}_3$ considered, 3-$SS$-$BB$ takes less time than $T$-$Rep$-$BB$ (which takes $3\ell=9027$ slots). Also, in Fig.~\ref{Sim2}, 3-$SS$-$BB$ takes more time than $T$-$Rep$-$BB$. Since $1500 < \zeta_1 (3) \times 3009 = 0.6286  \times 3009$ and $4000 > \zeta_2 (3) \times 3009 = 0.6622  \times 3009$, these observations are consistent with the result derived in Section~\ref{Boundary} that for  $\tilde{n}_1 \leq \zeta_1 (3) \times \ell = 0.6286 \ell$ (respectively, $\tilde{n}_1 \geq \zeta_2 (3) \times \ell= 0.6622 \ell$), 3-$SS$-$BB$ takes less (respectively, more) time than $T$-$Rep$-$BB$.

\begin{figure}
\centering

\begin{subfigure}{.5\textwidth}
\centering
\begin{adjustbox}{width = 1\columnwidth}
%
%
\begin{tikzpicture}

\begin{axis}[%
width=4.521in,
height=3.566in,
at={(0.758in,0.481in)},
scale only axis,
xmin=0.1,
xmax=0.5,
xtick={0.1, 0.2, 0.3, 0.4, 0.5},
yticklabel style = {font=\LARGE},
xticklabel style = {font=\LARGE},
xlabel style={font=\color{white!15!black}, font=\huge},
xlabel={$q$},
ymin=0,
ymax=100000,
ytick={0, 20000, 40000, 60000, 80000, 100000},
ylabel style={font=\color{white!15!black}, font=\huge},
ylabel={Number of slots required},
axis background/.style={fill=white},
title style={font=\color{white!15!black}, font=\huge},
title={Number of slots required v/s $q$},
legend style={font=\LARGE, at={(0.349,0.378)}, anchor=south west, legend cell align=left, align=left, draw=white!15!black}
]
\addplot [color=black, dotted, mark=x, mark options={solid, black}]
  table[row sep=crcr]{%
0.1	83839.072\\
0.2	88060.448\\
0.3	90989.056\\
0.4	93127.008\\
0.5	94903.712\\
};
\addlegendentry{3-SS}

\addplot [color=black, dashed, mark=diamond, mark options={solid, black}]
  table[row sep=crcr]{%
0.1	75220.9973333323\\
0.2	81376.9813333328\\
0.3	85885.7653333331\\
0.4	88403.1413333333\\
0.5	90612.6613333334\\
};
\addlegendentry{2-SS}

\addplot [color=black, dotted, mark=o, mark options={solid, black}]
  table[row sep=crcr]{%
0.1	10276.221\\
0.2	10306.002\\
0.3	10320.792\\
0.4	10332.835\\
0.5	10344.217\\
};
\addlegendentry{HSRC-1}

\addplot [color=black, dashed, mark=square, mark options={solid, black}]
  table[row sep=crcr]{%
0.1	7699.26966666666\\
0.2	7762.57566666666\\
0.3	7813.63066666667\\
0.4	7856.57766666667\\
0.5	7902.18966666667\\
};
\addlegendentry{HSRC-2}

\addplot [color=black]
  table[row sep=crcr]{%
0.1	12836\\
0.2	12836\\
0.3	12836\\
0.4	12836\\
0.5	12836\\
};
\addlegendentry{$T$ Repetitions of SRC$_S$}

\end{axis}
\end{tikzpicture}%
\end{adjustbox}
\caption{}
\label{Sim_q}
\end{subfigure}%
\begin{subfigure}{.5\textwidth}
\centering
\begin{adjustbox}{width = 1\columnwidth}
%
%
\begin{tikzpicture}

\begin{axis}[%
width=4.521in,
height=3.566in,
at={(0.758in,0.481in)},
scale only axis,
xmin=8,
xmax=256,
xtick={8, 32, 64, 128, 256},
yticklabel style = {font=\LARGE},
xticklabel style = {font=\LARGE},
xlabel style={font=\color{white!15!black}, font=\huge},
xlabel={$D$},
ymin=0,
ymax=100000,
ytick={0, 20000, 40000, 60000, 80000, 100000},
ylabel style={font=\color{white!15!black}, font=\huge},
ylabel={Number of slots required},
axis background/.style={fill=white},
title style={font=\color{white!15!black}, font=\huge},
title={Number of slots required v/s $D$},
legend style={font=\LARGE, at={(0.349,0.378)}, anchor=south west, legend cell align=left, align=left, draw=white!15!black}
]
\addplot [color=black, dotted, mark=x, mark options={solid, black}]
  table[row sep=crcr]{%
8	73515.104\\
16	75833.68\\
32	79800.592\\
64	84842.16\\
128	89381.616\\
256	91559.328\\
};
\addlegendentry{3-SS}

\addplot [color=black, dashed, mark=diamond, mark options={solid, black}]
  table[row sep=crcr]{%
8	56633.7653333328\\
16	63471.3493333325\\
32	69820.4533333323\\
64	76059.3653333324\\
128	82992.3733333329\\
256	91142.0373333334\\
};
\addlegendentry{2-SS}

\addplot [color=black, dotted, mark=o, mark options={solid, black}]
  table[row sep=crcr]{%
8	10175.906\\
16	10196.757\\
32	10231.345\\
64	10276.126\\
128	10315.688\\
256	10335.081\\
};
\addlegendentry{HSRC-1}

\addplot [color=black, dashed, mark=square, mark options={solid, black}]
  table[row sep=crcr]{%
8	7519.55966666666\\
16	7578.37866666666\\
32	7635.65666666666\\
64	7689.73666666666\\
128	7756.23666666666\\
256	7832.54866666667\\
};
\addlegendentry{HSRC-2}

\addplot [color=black]
  table[row sep=crcr]{%
8	12836\\
16	12836\\
32	12836\\
64	12836\\
128	12836\\
256	12836\\
};
\addlegendentry{$T$ Repetitions of SRC$_S$}

\end{axis}
\end{tikzpicture}%
\end{adjustbox}
\caption{}
\label{Sim_D}
\end{subfigure}
\caption{These plots show the average numbers of slots required by various estimation schemes. The following parameters are used: $T$ = 4, $\epsilon$ = 0.03, $D$ = 100  (in Fig.~\ref{Sim_q}) and $q$ = 0.15 (in Fig.~\ref{Sim_D}).}
\label{Sim6}
\end{figure}
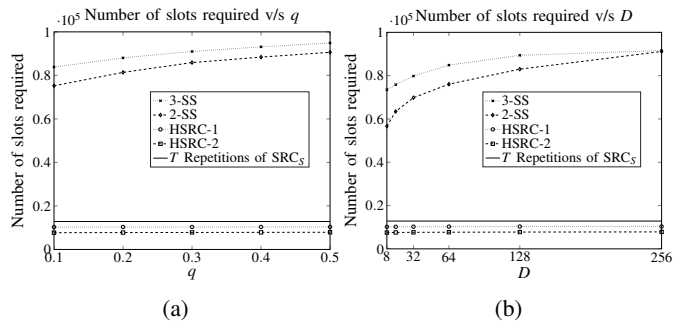

\begin{figure}
\centering
\begin{subfigure}{.5\textwidth}
\centering
\begin{adjustbox}{width = 1\columnwidth}
%
%
\begin{tikzpicture}

\begin{axis}[%
width=4.521in,
height=3.566in,
at={(0.758in,0.481in)},
scale only axis,
xmin=3,
xmax=8,
xtick={3, 4, 5, 6, 7, 8},
yticklabel style = {font=\LARGE},
xticklabel style = {font=\LARGE},
xlabel style={font=\color{white!15!black}, font=\huge},
xlabel={$T$},
ymin=0,
ymax=200000,
ytick={0, 50000, 100000, 150000, 200000},
ylabel style={font=\color{white!15!black}, font=\huge},
ylabel={Number of slots required},
axis background/.style={fill=white},
title style={font=\color{white!15!black}, font=\huge},
title={Number of slots required v/s $T$},
legend style={font=\Large, at={(0.02,0.648)}, anchor=south west, legend cell align=left, align=left, draw=white!15!black}
]
\addplot [color=black, dotted, mark=x, mark options={solid, black}]
  table[row sep=crcr]{%
3	61333.776\\
4	86326.912\\
5	113467.09\\
6	141442.22\\
7	165770.8\\
8	191623.89\\
};
\addlegendentry{3-SS}

\addplot [color=black, dashed, mark=diamond, mark options={solid, black}]
  table[row sep=crcr]{%
3	61333.776\\
4	80739.685\\
5	95219.52\\
6	123084.46\\
7	138735.51\\
8	169098.52\\
};
\addlegendentry{2-SS}

\addplot [color=black, dotted, mark=o, mark options={solid, black}]
  table[row sep=crcr]{%
3	7060.091\\
4	10290.073\\
5	13538.197\\
6	16792.843\\
7	20013.82\\
8	23250.999\\
};
\addlegendentry{HSRC-1}

\addplot [color=black, dashed, mark=square, mark options={solid, black}]
  table[row sep=crcr]{%
3	7060.091\\
4	7734.2117\\
5	8361.238\\
6	11616.269\\
7	12259.266\\
8	15530.54\\
};
\addlegendentry{HSRC-2}

\addplot [color=black]
  table[row sep=crcr]{%
3	9627\\
4	12836\\
5	16045\\
6	19254\\
7	22463\\
8	25672\\
};
\addlegendentry{$T$ Repetitions of SRC$_S$}

\end{axis}
\end{tikzpicture}%
\end{adjustbox}
\caption{}
\label{Sim_T}
\end{subfigure}%
\begin{subfigure}{.5\textwidth}
\centering
\begin{adjustbox}{width = 1\columnwidth}
%
%
\begin{tikzpicture}

\begin{axis}[%
width=4.521in,
height=3.566in,
at={(0.758in,0.481in)},
scale only axis,
xmin=0.01,
xmax=0.05,
xtick={0.01, 0.02, 0.03, 0.04, 0.05},
yticklabel style = {font=\LARGE},
xticklabel style = {font=\LARGE},
xlabel style={font=\color{white!15!black}, font=\huge},
xlabel={$\epsilon$},
ymin=0,
ymax=800000,
ylabel style={font=\color{white!15!black}, font=\huge},
ylabel={Number of slots required},
axis background/.style={fill=white},
title style={font=\color{white!15!black}, font=\huge},
title={Number of slots required v/s $\epsilon$},
legend style={font=\Large, at={(0.486,0.648)}, anchor=south west, legend cell align=left, align=left, draw=white!15!black}
]
\addplot [color=black, dotted, mark=x, mark options={solid, black}]
  table[row sep=crcr]{%
0.01	760309.008\\
0.02	192495.205\\
0.03	86080.4\\
0.04	48972.614\\
0.05	31669.482\\
};
\addlegendentry{3-SS}

\addplot [color=black, dashed, mark=diamond, mark options={solid, black}]
  table[row sep=crcr]{%
0.01	699983.249333327\\
0.02	177247.617333332\\
0.03	79336.7253333326\\
0.04	45181.0246666663\\
0.05	29186.5249999997\\
};
\addlegendentry{2-SS}

\addplot [color=black, dotted, mark=o, mark options={solid, black}]
  table[row sep=crcr]{%
0.01	84914.271\\
0.02	21780.714\\
0.03	10288.232\\
0.04	6060.632\\
0.05	4164.245\\
};
\addlegendentry{HSRC-1}

\addplot [color=black, dashed, mark=square, mark options={solid, black}]
  table[row sep=crcr]{%
0.01	62707.843\\
0.02	16188.5163333333\\
0.03	7720.47566666666\\
0.04	4608.94566666666\\
0.05	3208.25899999999\\
};
\addlegendentry{HSRC-2}

\addplot [color=black]
  table[row sep=crcr]{%
0.01	107100\\
0.02	27352\\
0.03	12836\\
0.04	7496\\
0.05	5100\\
};
\addlegendentry{$T$ Repetitions of SRC$_S$}

\end{axis}
\end{tikzpicture}%
\end{adjustbox}
\caption{}
\label{Sim_eps}
\end{subfigure}
\caption{These plots show the average number of slots required by various estimation schemes. The following parameters are used: $D$ = 100, $q$ = 0.15, $\epsilon$ = 0.03 (in Fig.~\ref{Sim_T}) and $T$ = 4  (in Fig.~\ref{Sim_eps}).}
\label{Sim7}
\end{figure}
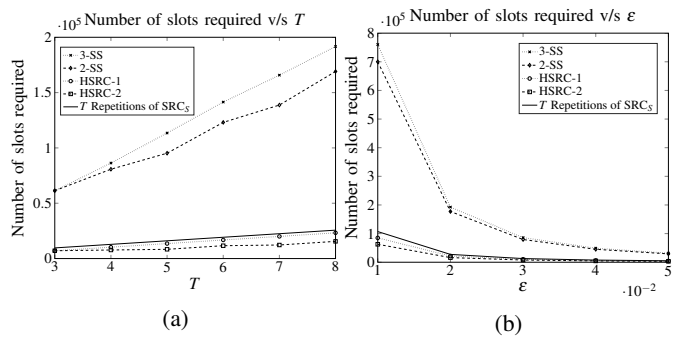

Now, we compare the performances of the proposed schemes, viz., HSRC-1 and HSRC-2, with those of the scheme in which the  SRC$_S$ protocol proposed in~\cite{zhou2016understanding} is separately executed $T$ times to estimate the active node cardinality of each node type, and  the 3-$SS$ and 2-$SS$ schemes proposed in our prior work~\cite{kadam2017fast},~\cite{TechReport2018}. For a fair comparison, all the schemes are executed as many times as is required to achieve the same accuracy level $\epsilon = 0.03$.
In phase 2 of HSRC-1, we use the method (either 3-$SS$-$BB$ or $T$-$Rep$-$BB$) that requires fewer slots. Similarly, in phase 2 of HSRC-2, we use the method (either 2-$SS$-$BB$ or $T$-$Rep$-$BB$) that requires fewer slots.    
Fig.~\ref{Sim_q} (respectively, Fig.~\ref{Sim_D}) shows a plot of the number of slots required by various estimation schemes versus $q$ (respectively, $D$). Fig.~\ref{Sim_q} and Fig.~\ref{Sim_D} show that the proposed schemes significantly outperform 3-$SS$ and 2-$SS$, and also outperform the scheme in which the  SRC$_S$ protocol is executed $T$ times. In Fig.~\ref{Sim_q}, HSRC-2 (respectively, HSRC-1) outperforms the $T$ repetitions of SRC$_S$ protocol by 39.18\% (respectively, 19.63\%) on average. Also, in Fig.~\ref{Sim_D}, HSRC-2 (respectively, HSRC-1) outperforms the $T$ repetitions of SRC$_S$ protocol by 40.25\% (respectively, 20.11\%) on average.   Among the proposed schemes, HSRC-2 performs better than HSRC-1. Since the SRC$_S$ protocol has been shown to significantly outperform the LoF based protocol in~\cite{zhou2016understanding}, the $T$ repetitions of SRC$_S$ protocol performs better than 3-$SS$ and 2-$SS$~\cite{kadam2017fast},~\cite{TechReport2018}, which are both designed by extending the LoF based estimation scheme to heterogeneous networks.

Fig.~\ref{Sim_T} (respectively, Fig.~\ref{Sim_eps}) shows a plot of the number of slots required by various estimation schemes versus $T$ (respectively, $\epsilon$). Figs.~\ref{Sim_T} and~\ref{Sim_eps} both show trends that are similar to those in Figs.~\ref{Sim_q} and~\ref{Sim_D}. In Fig.~\ref{Sim_T}, HSRC-2 (respectively, HSRC-1) outperforms the $T$ repetitions of SRC$_S$ protocol by 33.29\% (respectively, 15.86\%) on average. Also, in Fig.~\ref{Sim_eps}, HSRC-2 (respectively, HSRC-1) outperforms the $T$ repetitions of SRC$_S$ protocol by 39.54\% (respectively, 19.69\%) on average.

\section{Conclusions}\label{Conc}
We designed two schemes, viz., HSRC-1 and HSRC-2, for rapidly obtaining separate estimates of the number of active nodes of each type in a heterogeneous M2M network with $T$ types of nodes, where $T \geq 2$ is an arbitrary integer. Our schemes consist of two phases; we analytically derived a condition that can be used to decide as to which of two possible approaches should be used in phase 2 of HSRC-1 to minimize its execution time. The expected number of slots required by HSRC-1 to execute  and the expected energy consumption of a node under  HSRC-1 were analysed. Using simulations, we showed that our proposed schemes, HSRC-1 and HSRC-2, require significantly fewer time slots to execute compared to estimation schemes designed for heterogeneous networks in prior work, viz., 3-$SS$ and 2-$SS$, and also compared to separately executing the underlying estimation protocol, SRC$_S$~\cite{zhou2016understanding}, for homogeneous networks  $T$ times, even though all these schemes obtain estimates with the same accuracy.  

\bibliography{BibFiles} 
\bibliographystyle{ieeetr}

\appendix
\begin{IEEEproof}[Proof of Proposition~\ref{Case1_Cond}]
Using~\eqref{EQ:pb},~\eqref{eqQ1}--\eqref{eqvnb}, we get:
\begin{align}
Q_1 &= 1 - \Big(1 - \frac{1}{\ell}\Big)^{{\tilde{n}}_{1}} - \frac{{\tilde{n}}_{1}}\ell\Big(1 - \frac{1}\ell\Big)^{{\tilde{n}}_{1}-1}, \label{EQ:Q1} \\
Q_2 &=  \frac{{\tilde{n}}_{1}}{\ell}\Big(1 - \frac{1}\ell\Big)^{{\tilde{n}}_{1}-1}\Big(1 - {e}^{-1.6}\Big)^{T-1} \nonumber \\ 
	&= (0.7981)^{T-1} \times \frac{{\tilde{n}}_{1}}{\ell}\Big(1 - \frac{1}\ell\Big)^{{\tilde{n}}_{1}-1}, \label{EQ:Q2} \\
Q_3 &=  \Big(1 - \frac{1}\ell\Big)^{{\tilde{n}}_{1}}\Big(1 - 2.6 {e}^{-1.6}\Big)^{T-1} \nonumber \\ 
	&=(0.4751)^{T-1} \times \Big(1 - \frac{1}\ell\Big)^{{\tilde{n}}_{1}}. \label{EQ:Q3}
\end{align}
Now consider the LHS of \eqref{EQ:cond2} (which is the same as that of \eqref{EQ:cond3}). By \eqref{EQ:EK} and \eqref{EQ:ER}: 
\begin{align}
\label{EQ:Q_all}
&(1 + 1/S_W) E(K_T) +  (T-1)E(R_T)  \nonumber  \\
&= \ell (1 + 1/S_W) (Q_1 + Q_2 + Q_3) + \ell (T - 1) Q_1 \nonumber \\
&= \ell \Big[(T + 1/S_W) Q_1 + (1 + 1/S_W) (Q_2 + Q_3)\Big].
\end{align}
By substituting~\eqref{EQ:Q1},~\eqref{EQ:Q2}, and~\eqref{EQ:Q3} into~\eqref{EQ:Q_all}, we get:
\begin{align}
\label{EQ:NewCond}
&(1 + 1/S_W) E(K_T) +  (T-1)E(R_T) \nonumber \\
& =   \ell \Big[(T + 1/S_W) \Big(1 - \Big(1 - \frac{1}{\ell}\Big)^{{\tilde{n}}_{1}} - \frac{{\tilde{n}}_{1}}\ell\Big(1 - \frac{1}\ell\Big)^{{\tilde{n}}_{1}-1}\Big) \nonumber \\
& + (1 + 1/S_W) \Big((0.7981)^{T-1} \times \frac{{\tilde{n}}_{1}}{\ell}\Big(1 - \frac{1}{\ell}\Big)^{{\tilde{n}}_{1}-1} \nonumber \\ 
&+ (0.4751)^{T-1} \times \Big(1 - \frac{1}\ell\Big)^{{\tilde{n}}_{1}}\Big) \Big] \nonumber \\
& = \ell \Big[(T + 1/S_W)  -  \Big(1 - \frac{1}\ell\Big)^{{\tilde{n}}_{1}} \Big\{  (T + 1/S_W) \nonumber \\
& - (1 + 1/S_W) (0.4751)^{T-1} \Big\}   - \frac{{\tilde{n}}_{1}}\ell\Big(1 - \frac{1}\ell\Big)^{{\tilde{n}}_{1}-1} \nonumber \\
& \Big\{  (T + 1/S_W) - (1 + 1/S_W) (0.7981)^{T-1}\Big\} \Big].
\end{align}
Substituting from \eqref{EQ:NewCond} into \eqref{EQ:cond2} and simplifying, we get: 
\begin{align}
\label{EQ:NewCond4}
 & \Big(1 - \frac{1}\ell\Big)^{{\tilde{n}}_{1}} \Big\{  (1 + T S_W) - (1 + S_W) (0.4751)^{T-1} \Big\}  + \nonumber \\
 & \frac{{\tilde{n}}_{1}}\ell\Big(1 - \frac{1}\ell\Big)^{{\tilde{n}}_{1}-1} \Big\{  (1 + T S_W) - (1 + S_W) (0.7981)^{T-1}\Big\} \nonumber \\
 & \ge (T-1) S_W + 2 + 2S_W/\ell.
\end{align}
Similarly, substituting from \eqref{EQ:NewCond} into \eqref{EQ:cond3} and simplifying, we get:  
\begin{align}
\label{EQ:NewCond4a}
 & \Big(1 - \frac{1}\ell\Big)^{{\tilde{n}}_{1}} \Big\{  (1 + T S_W) - (1 + S_W) (0.4751)^{T-1} \Big\}  + \nonumber \\
 & \frac{{\tilde{n}}_{1}}\ell\Big(1 - \frac{1}\ell\Big)^{{\tilde{n}}_{1}-1} \Big\{  (1 + T S_W) - (1 + S_W) (0.7981)^{T-1}\Big\} \nonumber \\
 & \ge (T-1) S_W + 2.
\end{align}

Now, let $F_1(T, S_W) = (1 + T S_W) - (1 + S_W) (0.4751)^{T-1}$ and $F_2(T, S_W) = (1 + T S_W) - (1 + S_W) (0.7981)^{T-1}$. So,~\eqref{EQ:NewCond4} simplifies to: 
\begin{align}
\label{EQ:NewCond5}
& F_1 (T, S_W) \Big(1 - \frac{1}\ell\Big)^{{\tilde{n}}_{1}}  + F_2 (T, S_W) \frac{{\tilde{n}}_{1}}\ell\Big(1 - \frac{1}\ell\Big)^{{\tilde{n}}_{1}-1}  \nonumber \\
 & \ge (T-1) S_W + 2 + 2S_W/\ell.
\end{align}
Similarly,~\eqref{EQ:NewCond4a} simplifies to: 
\begin{align}
\label{EQ:NewCond5a}
& F_1 (T, S_W) \Big(1 - \frac{1}\ell\Big)^{{\tilde{n}}_{1}}  + F_2 (T, S_W) \frac{{\tilde{n}}_{1}}\ell\Big(1 - \frac{1}\ell\Big)^{{\tilde{n}}_{1}-1}  \nonumber \\
 & \ge (T-1) S_W + 2.
\end{align}
Since $S_W = 6$, $F_1(T, 6) = (1 + 6T) - 7(0.4751)^{T-1} = G_1(T)$ and $F_2(T, 6) = (1 + 6T) - 7(0.7981)^{T-1} = G_2(T)$. With these substitutions, \eqref{EQ:NewCond5} (respectively, \eqref{EQ:NewCond5a}) simplifies to \eqref{EQ:NewCond6} (respectively, \eqref{EQ:NewCond6a}).
 \end{IEEEproof}
 
 \begin{IEEEproof}[Proof of Proposition~\ref{PN:Case1}]
Consider:
\begin{eqnarray}
& G_1(T) \Big(1 - \frac{1}{\ell}\Big)^{{\tilde{n}}_{1}} + G_2(T) \frac{{\tilde{n}}_{1}}{\ell}\Big(1 - \frac{1}{\ell}\Big)^{{\tilde{n}}_{1}-1}  \nonumber \\& =
	G_1(T)  \Big(1 - \frac{1}{\ell}\Big)^{{\tilde{n}}_{1}} +
	G_2(T) \frac{\frac{{\tilde{n}}_{1}}{\ell}\Big(1 - \frac{1}{\ell}\Big)^{{\tilde{n}}_{1}}}{1-\frac{1}{\ell}}   \nonumber  \\&
	\geq G_1(T) \Big(1 - \frac{1}{\ell}\Big)^{{\tilde{n}}_{1}} + G_2(T) \frac{{\tilde{n}}_{1}}{\ell}\Big(1 - \frac{1}{\ell}\Big)^{{\tilde{n}}_{1}}.  
	\label{ntilde1_T}	
\end{eqnarray}
Let ${\tilde{n}}_{1}=x\ell$. Then the quantity in (\ref{ntilde1_T}) equals: 
	 $(1 - \frac{1}{\ell})^{x\ell}(G_1(T)+  xG_2(T))$. Now, it can be easily shown that the function $g(\ell)=(1-\frac{1}{\ell})^\ell$ is increasing in $\ell$. Since $\ell\geq100$, $g(\ell)\geq g(100) = 0.366$. Hence, the quantity in (\ref{ntilde1_T}): $ \geq (0.366)^{x}(G_1(T)+  xG_2(T))= f(x, T)$ (which is defined in Section~\ref{Case1}).

	Now, by the definition of $\zeta_1(T)$,  $f(\zeta_1(T), T) \ge 6T - 3.88 \ge 6T - 4 +12/\ell$ (since $\ell \geq100$). Hence, for $x \leq \zeta_1(T)$, $f(x, T)\geq f(\zeta_1(T), T)$.\footnote{This holds since $f(x,T)$ is a decreasing function for $x>0$ (see Section~\ref{Boundary}).} It follows that the quantity in  (\ref{ntilde1_T}) is $\geq f(\zeta_1(T), T)$ for $x \leq \zeta_1(T)$, or equivalently, ${\tilde{n}}_{1} \leq \zeta_1(T)\ell$. Hence, inequality (\ref{EQ:NewCond6}) holds for ${\tilde{n}}_{1} \leq \zeta_1(T)\ell$ and $\ell \geq100$.
	
Next, consider: 
\begin{align}
	& G_1(T) \left(1 - \frac{1}{\ell}\right)^{{\tilde{n}}_{1}} + G_2(T) \frac{{\tilde{n}}_{1}}{\ell}\left(1 - \frac{1}{\ell}\right)^{{\tilde{n}}_{1}-1}  \nonumber \\&
	\leq G_1(T) \left(1 - \frac{1}{\ell}\right)^{{\tilde{n}}_{1}} + \frac{G_2(T)}{0.99} \frac{{\tilde{n}}_{1}}{\ell}\left(1 - \frac{1}{\ell}\right)^{{\tilde{n}}_{1}}\ \mbox{(since $\ell\geq 100$)} \  \nonumber \\&
= \left(1 - \frac{1}{\ell}\right)^{x\ell}\left(G_1(T) + x\frac{G_2(T)}{0.99}\right) \mbox{  (using ${\tilde{n}}_{1}=x\ell$) }.
\label{ntilde2_T}	 
\end{align}
 Now, $g(\ell)=\left(1-\frac{1}{\ell}\right)^\ell < g(\infty)=e^{-1}=0.3679$. Hence, the quantity in (\ref{ntilde2_T}): $< (0.3679)^{x}\left(G_1(T) + x\frac{G_2(T)}{0.99}\right)= f_1(x,T)$ (which is defined in Section~\ref{Case1}). 
 Now, by definition of $\zeta_2(T)$, $f_1(\zeta_2(T), T) < 6T -4$.\footnote{This holds since $f_1(x,T)$ is a decreasing function for $x>0$ (see Section~\ref{Boundary}).} Hence, for $x \geq \zeta_2(T)$, $f_1(x, T) < f_1(\zeta_2(T), T) < 6T -4$. Hence, inequality (\ref{EQ:NewCond6a}) does not hold when $\zeta_2(T) \ell \leq {\tilde{n}}_1 <1.6\ell$.    
 \end{IEEEproof}

\begin{IEEEproof}[Proof of Proposition~\ref{Case2_Cond}]
Using~\eqref{EQ:pb},~\eqref{eqQ1}--\eqref{eqvnb},  we get:
\begin{align}
Q_1  	&= 1 - \left(1 - \frac{1.6}{\tilde{n}_{1}}\right)^{{\tilde{n}}_{1}} -1.6\left(1 - \frac{1.6}{\tilde{n}_{1}}\right)^{{\tilde{n}}_{1}-1}, \label{EQ:Q1_2} \\
Q_2 &=  1.6\left(1 - \frac{1.6}{\tilde{n}_{1}}\right)^{{\tilde{n}}_{1}-1}\left(1 - {e}^{-1.6}\right)^{T-1} \nonumber \\ 
	&= (0.7981)^{T-1} \times 1.6\left(1 - \frac{1.6}{\tilde{n}_{1}}\right)^{{\tilde{n}}_{1}-1}, \label{EQ:Q2_2} \\
Q_3 &= \left(1 - \frac{1.6}{\tilde{n}_{1}}\right)^{{\tilde{n}}_{1}}\left(1 - 2.6 {e}^{-1.6}\right)^{T-1} \nonumber \\ 
	&=(0.4751)^{T-1} \times \left(1 - \frac{1.6}{\tilde{n}_{1}}\right)^{{\tilde{n}}_{1}}. \label{EQ:Q3_2}
\end{align}
Now, by following a procedure similar to that in \eqref{EQ:Q_all}, \eqref{EQ:NewCond}, \eqref{EQ:NewCond4a}, \eqref{EQ:NewCond5a} and replacing $\left(1 - \frac{1}\ell\right)^{{\tilde{n}}_{1}}$ with $\left(1 - \frac{1.6}{\tilde{n}_{1}}\right)^{{\tilde{n}}_{1}}$ and $\frac{{\tilde{n}}_{1}}{\ell}\left(1 - \frac{1}\ell\right)^{{\tilde{n}}_{1}-1}$ with $1.6\left(1 - \frac{1.6}{\tilde{n}_{1}}\right)^{{\tilde{n}}_{1}-1}$, we get that a necessary condition for~\eqref{EQ:cond1} to hold is: 
\begin{align}
\label{EQ:NewCond7}
 & F_1 (T, S_W) \left(1 - \frac{1.6}{\tilde{n}_{1}}\right)^{{\tilde{n}}_{1}}  + F_2 (T, S_W) 1.6\left(1 - \frac{1.6}{\tilde{n}_{1}}\right)^{{\tilde{n}}_{1}-1}  \nonumber \\
 & \ge (T-1) S_W + 2.
\end{align}
When $S_W = 6$,~\eqref{EQ:NewCond7} simplifies to~\eqref{EQ:NewCond8}.
 \end{IEEEproof}

\begin{IEEEproof}[Proof of Proposition~\ref{PN:Case2}]
It is easy to show that $\left(1 - \frac{1.6}{\tilde{n}_1}\right)^{{\tilde{n}}_{1}}$ is increasing in $\tilde{n}_1$ and its maximum value is $e^{-1.6} = 0.202$ at $\tilde{n}_1$=$\infty$. Consider:
 \begin{eqnarray*}
 & & G_1(T) \left(1 - \frac{1.6}{\tilde{n}_{1}}\right)^{{\tilde{n}}_{1}}   +  G_2(T) 1.6\left(1 - \frac{1.6}{\tilde{n}_{1}}\right)^{{\tilde{n}}_{1}-1} \\
 & = &\left(1 - \frac{1.6}{\tilde{n}_1}\right)^{{\tilde{n}}_{1}}\left( G_1(T) + \frac{1.6}{(1 - \frac{1.6}{\tilde{n}_1})} G_2(T)\right) \\
 & < & e^{-1.6} \left( G_1(T) + \frac{1.6}{(1 - \frac{1.6}{\tilde{n}_1})} G_2(T)\right)\\
 & \leq & e^{-1.6} \left( G_1(T) + \frac{1.6}{0.99} G_2(T) \right) \\
 & &\mbox{  (since } \tilde{n}_1 \geq 1.6\ell \mbox{ and } \ell \geq 100) \\
 & = & 0.202 ( G_1(T) + 1.616 G_2(T))\\
 & = & 0.202  \Big((1 + 6T) - 7(0.4751)^{T-1} \\
 & & + 1.616\times(1 + 6T) - 7\times1.616\times(0.7981)^{T-1} \Big)\\
 & \le & 3.1706T + 0.5284 \mbox{  (since the minimum values of } \\
 & & (0.4751)^{T-1} \mbox{ and } (0.7981)^{T-1} \mbox{ are } 0)\\
 & < & 6T - 4 \hspace{10mm} (\mbox{since } T \ge 2).
 \end{eqnarray*}
The result follows. 
\end{IEEEproof}

\end{document}